\newcommand*{\ARXIV}{}
\ifdefined\ARXIV
\else
\fi

\ifdefined\ARXIV
\documentclass{article}
\usepackage{arxiv}

\else
\documentclass[journal]{vgtc}                
\fi

\ifpdf
  \pdfoutput=1\relax                   
  \usepackage{graphicx}                
  \DeclareGraphicsExtensions{.pdf,.png,.jpg,.jpeg} 
\else
  \usepackage{graphicx}                
  \DeclareGraphicsExtensions{.eps}     
\fi%

\graphicspath{{figures/}{pictures/}{images/}{./}} 

\usepackage{microtype}                 
\PassOptionsToPackage{warn}{textcomp}  
\usepackage{textcomp}                  
\usepackage{mathptmx}                  
\usepackage{times}                     
\usepackage{cite}                      
\usepackage{tabu}                      
\usepackage{booktabs}                  
\usepackage{url}                  
\usepackage{enumitem}

\ifdefined\ARXIV

\else

\onlineid{1168}

\vgtccategory{Research}
\vgtcpapertype{evaluation}

\fi

\title{Directions for 3D User Interface Research from Consumer VR Games}

\ifdefined\ARXIV

\author{Anthony Steed\\
Department of Computer Science\\University College London\\
\texttt{A.Steed@ucl.ac.uk}
\And
Tuukka M. Takala\\
Aalto University \\ Waseda University\\
\texttt{tuukka.takala@aalto.fi}
\And
Daniel Archer\\
Department of Computer Science\\University College London\\
\texttt{D.Archer@ucl.ac.uk}
\And
Wallace Lages\\
School of Visual Arts\\
Virginia Tech\\
\texttt{wlages@vt.edu}
\And
Robert W. Lindeman\\
HIT Lab NZ\\
University of Canterbury\\
\texttt{gogo@hitlabnz.org}
}

\else

\author{Anthony Steed \textit{Member, IEEE}, Tuukka M. Takala,  Daniel Archer, Wallace Lages \\ and Robert W. Lindeman \textit{Senior Member, IEEE}}
\authorfooter{
\item
Anthony Steed is with Department of Computer Science, University College London. E-mail: A.Steed@ucl.ac.uk.
\item
Tuukka M. Takala is with Aalto University and Waseda University. E-mail: tuukka.takala@aalto.fi.
\item
Daniel Archer is with Department of Computer Science, University College London. E-mail: D.Archer@ucl.ac.uk.
\item
Wallace Lages is with School of Visual Arts, Virginia Tech. E-mail: wlages@vt.edu.
\item
Robert W. Lindeman is with the HIT Lab NZ at the University of Canterbury. E-mail: gogo@hitlabnz.org.}

\shortauthortitle{Steed \MakeLowercase{\textit{et al.}}: Directions for 3D User Interface Research from Consumer VR Games}

\fi

\ifdefined\ARXIV
\begin{document}
\maketitle

\else
\fi


\ifdefined\ARXIV
\begin{abstract}
\else
\abstract{
\fi

With the continuing development of affordable immersive virtual reality (VR) systems, there is now a growing market for consumer content. The current form of consumer systems is not dissimilar to the lab-based VR systems of the past 30 years: the primary input mechanism is a head-tracked display and one or two tracked hands with buttons and joysticks on hand-held controllers. Over those 30 years, a very diverse academic literature has emerged that covers design and ergonomics of 3D user interfaces (3DUIs). However, the growing consumer market has engaged a very broad range of creatives that have built a very diverse set of designs. Sometimes these designs adopt findings from the academic literature, but other times they experiment with completely novel or counter-intuitive mechanisms. In this paper and its online adjunct, we report on novel 3DUI design patterns that are interesting from both design and research perspectives: they are highly novel, potentially broadly re-usable and/or suggest interesting avenues for evaluation. The supplemental material, which is a living document, is a crowd-sourced repository of interesting patterns. This paper is a curated snapshot of those patterns that were considered to be the most fruitful for further elaboration. 

\ifdefined\ARXIV
\end{abstract}
\else
} 
\fi


\ifdefined\ARXIV
\else



\CCScatlist{ 
 \CCScat{K.6.1}{Management of Computing and Information Systems}%
{Project and People Management}{Life Cycle};
 \CCScat{K.7.m}{The Computing Profession}{Miscellaneous}{Ethics}
}




\vgtcinsertpkg
\fi


\ifdefined\ARXIV

\section{Introduction}

\else

\begin{document}

\firstsection{Introduction}

\maketitle

\fi

In the 2010s, the virtual reality (VR) market moved at remarkable speed from a high-end professional and academic market to a consumer market. At the time of writing, there is still a healthy rate of new devices coming to market. There is a move towards cheaper devices that provide full immersion, as well as more expensive devices that push out the fidelity limits.  Alongside the consumer devices, there is a plethora of consumer applications: many games, dramas, documentaries, creative applications, health applications and so on. There are now thousands of applications that can be experienced alongside millions of panoramic videos. 

We note that the market for VR devices did involve a temporary technical step backwards. Early high-end immersive VR systems, such as the VPL Research RB2, Division Provision 100 and Virtuality 1000CS, provided six-degree-of-freedom (6DOF, 3 degrees of displacement and 3 degrees of rotation) tracking for the head and 6DOF tracking for one or two hand-held controllers. The early consumer market started with devices supporting only 3DOF head tracking (e.g., Oculus DK1, Google Cardboard) and no hand controllers. Today, the consumer market covers devices with a wide range of configurations which has implications we will cover later. However, on the flip side, aside from cost, the current range of devices is light, sometimes untethered and more easily deployable than those previous lab-only systems. Now, more systems are moving beyond hand-held controllers to bare-hand interaction, a capability that usually could only be prototyped with encumbering devices in labs in the 1990s and 2000s. 

This creates a situation where 3D user interface (3DUI) research that might be applicable to modern consumer VR systems stretches back to the early 1990s. This span of time creates challenges for current researchers and developers who are looking for rules of thumb, design guidelines or specific techniques to use with consumer VR systems. 3DUI research can also be found in a variety of academic and industrial venues. There is a separate problem of applying research results from older systems that had, for example, lower screen resolutions or higher latencies, to modern systems; sometimes it is not clear if results generalise and/or whether experiments need to be re-run to inform design decisions. All of this has led to some re-invention of the wheel in modern systems. However, there are underlying principles that still shine through, such as the classic separation of selection, manipulation, locomotion, system control and symbolic input\cite{laviola_jr_3d_2017}, the role of the body of the user in the system \cite{slater_body_1994,mine_moving_1997}, reality-based interaction paradigms \cite{jacob_reality-based_2008} and design processes specific for immersive systems \cite{jerald_vr_2015}.

In this paper, we take a look at the current range of designs for experiences in consumer VR and ask what generalisable techniques might be identified that expand the repertoire of 3DUI. We are looking for design patterns that might be quite general purpose or at least applicable to a large subset of systems, and that push 3DUI in new directions.  

The first aim is to identify patterns that deserve further study outside of a specific design context. That is, they might be generalisable, but we need a further elaboration of the positive and negative attributes. The second aim is to foster collaboration between research labs and designers, by highlighting the tensions between user research and design exploration. For example, as the user community grows, we might need to establish better guidelines for designers, highlight best practices and support greater accessibility. 

Because there are now many thousands of experiences available for consumer VR, our approach has not been to systematically review all experiences or to sample popular experiences. We acknowledge that novelty might be found in smaller titles or even demonstrations. Thus, we have crowd-sourced suggestions from a variety of email lists, Slack channels and other social media.

This paper is not an attempt to generate guidelines for future designers nor a rule book to follow. It simply highlights some areas of 3DUI design that the authors collectively feel deserve more attention. We do, however, discuss the differences between motivation and constraints that consumer VR experience designers have taken on board, compared to traditional academics.
 
In Section \ref{sec:framing}, we give a broader context to 3DUI research and outline some specific concerns about 3DUI that are highlighted in the consumer environment. Section \ref{sec:taxonomy} then discusses the taxonomy of techniques we present, how the examples were chosen and a description of the goals. The following four sections introduce areas of classic 3DUI concern, Selection (Section \ref{sec:selection}), Manipulation (Section \ref{sec:manipulation}), Locomotion (Section \ref{sec:locomotion}) and System Control (Section  \ref{sec:systemcontrol}). 
We add a fifth area of Miscellaneous Techniques (Section \ref{sec:misc}) that mostly deals with systems or specific situations of the user and their comfort in the system. This is motivated by the framing issues we raise in Section \ref{sec:framing}. We then conclude and present ongoing activities in this area (Section \ref{sec:conclusion}).

\section{Background and Framing}
\label{sec:framing}
\label{sec:background}

In this section, we consider how recent work on consumer VR fits within the frameworks and trends of 3DUI. We would emphasise that this is not to criticise either the commercial or academic work, but rather to highlight the slightly different concerns that each has. While in general, design of 3DUI in any context prioritises effectiveness and ease of use, as would any user interface, consumer VR experiences are obviously designed with interesting game play or user engagement as a priority, rather than simply the effectiveness of a technique as a goal unto itself. For example, the design of a game experience places many different constraints on the choice and style of interaction (e.g., visual representation, simple things such as gravity). The first implication of this is that in our later review we have focused on techniques that we think are generalisable rather than working for a very specific game because of the context. The second implication is that techniques work together, but relatively little academic work has looked at the complementarity of techniques, whereas a game experience has to consider the whole design. Thus, we leave consideration of the ways in which design of a coherent experience might affect specific choices of techniques to future work, and here focus on some specific techniques and their differences to what we might call ``established practice.''

\subsection{Narrow Interpretation of 3DUI}

A key observation is that, while there are a variety of consumer VR systems, they form a rather narrow subset of what has previously been considered within the scope of 3DUI~\cite{laviola_jr_3d_2017}. First, the consumer systems that have become popular over the past five years are HMD-based, whereas academic work has focused on a variety of VR systems including immersive projection  systems~\cite{cruz-neira_surround-screen_1993}, desktop or table systems~\cite{agrawala_two-user_1997} or novel 6DOF input devices~\cite{froehlich_globefish_2006}. It might be fair to say that some of that work was done partly because HMDs of the era were not as effective. However, such devices have found niche industrial applications over the past 20 or so years, support different modes of working including working for long periods, and support multiple users. There is, for example, still a lot of ongoing work on multi-user, but unencumbered displays~\cite{munoz-arango_multi-user_2019}. 


A second observation is that consumer systems tend to have very similar configurations for hand controllers. While we will discuss consumer systems without 6DOF tracking for both hands, at the time of writing, there is one dominant form of control for most systems: the user holds two controllers, one in each hand, and these are multipurpose controllers with, usually, a joystick or touchpad, a trigger and several other buttons. Within the broader 3DUI context, there is a lot of work on various forms of hand-controllers with force feedback (e.g., \cite{choi_claw_2018}) or controllers that change shape (e.g., \cite{krekhov_self-transforming_2017}). While there are various projects that aim to make new consumer devices such as gloves, advanced controllers, or task-specific controllers, no one device has become very common. 

Third, a few aspects of 3DUI are out of scope for this review. While there are options for adding full-body motion capture, these are still very much in the domain of the professional user or serious hobbyist, with relatively few experiences supporting highly articulated avatars. However, we do see this as an emerging area of work. Applications such as \textit{\textbf{VRChat}}~\footnote{Here and throughout, if we refer to an application, it will be in bold italics and listed in a section before the references for easy cross-referencing.} suggest that this will become more important going forward, as social VR becomes more prevalent.

The relative narrowness of the 3DUI currently being explored in industry does mean that content developed for one consumer system is relatively easy to port to other systems, requiring perhaps something as simple as a new button mapping. Standards such as OpenXR~\cite{krohnos_group_openxr_2016} will possibly allow developers of new devices to back-support older content. However, we expect that there will be a need for new standards that support a broader range of 3DUI for future VR systems.  

\subsection{Dealing with the Real Environment}
\label{sec:dealing}

It is an over-generalisation, but much 3DUI work done in academia to date has made assumptions about the interface being positioned in a relatively large open space, be it a tracked space in a lab or an immersive projection system. These spaces are probably larger than most consumer VR use cases. Further, consumer systems are often used solo and thus without a helpful lab assistant or student managing cables, moving obstacles and shepherding the user away from collisions. Thus, developers of consumer equipment and experiences have had to plan to deal with the real world as it is found.

This is accommodated, at one level, by the now-common use of chaperone systems that are provided by the run-time platform software. These allow the user to delineate their own free space or, in some cases, switch to a fixed position. Experiences thus need to adapt to where the user is within this self-defined space, and keep actions of the user inside the chaperone as much as possible; the user may injure themselves or cause other damage if they stray outside the space without warning. Of course, the free space might change because of temporary objects, or mobile objects (the so-called ``cat problem,'' where the family cat comes and sits in the free space without being detected by the system or user\cite{steed_vrs_2020}). 

\subsection{Accessibility}
\label{sec:access}

The issue of broad accessibility is one that has challenged VR content developers over the past few years. Again, it is perhaps reasonable to claim that most academic and prior studies were done with able-bodied users. There has also been a strong bias towards males~\cite{peck_mind_2020}. 
One exception is the rehabilitation community, but the focus there has tended to be on the creation of content to facilitate, encourage or monitor exercise (e.g.,~\cite{adamovich_sensorimotor_2009, rose_immersion_2018}). The challenge is that the default assumption of many researchers is of an able user, who will stand or sit, and actively move their arms to pick and place objects, etc. Some interfaces require two hands, while others require the user to stand to reach objects, and these immediately exclude users who cannot carry out such actions. Whereas, in the real world, various aids or adaptations might be made, the consumer VR experience is currently quite inflexible. 

Accessibility is of course a societal need, so we should expect more emphasis, and eventually standards, on accessibility to VR content. Some direct interventions have been made, such as the Microsoft SeeingVR toolkit that shows how visual aids might be added to applications by the developer or the user~\cite{zhao_seeingvr_2019}, or \textbf{\textit{WalkinVR}}, a plugin to SteamVR that facilitates modifications to enable users with limited or uncontrolled motion to use consumer VR systems.

There are broader angles to accessibility, such as dealing with all body shapes and sizes, especially if the environment is suitable for children, supporting people with larger hair arrangements~\cite{mboya_oculus_2020}, or simply dealing with a desire to sit down when the platform supports this. Environment developers, however, normally anticipate the user will be standing. This issue is linked to the issues in the previous section, because users may prefer to sit to feel safer, or because their environment has more reachable space if the user sits.

\subsection{Diegetic Interfaces}
\label{sec:diegetic}

\begin{figure}
    \centering
    \includegraphics[width=0.5\linewidth]{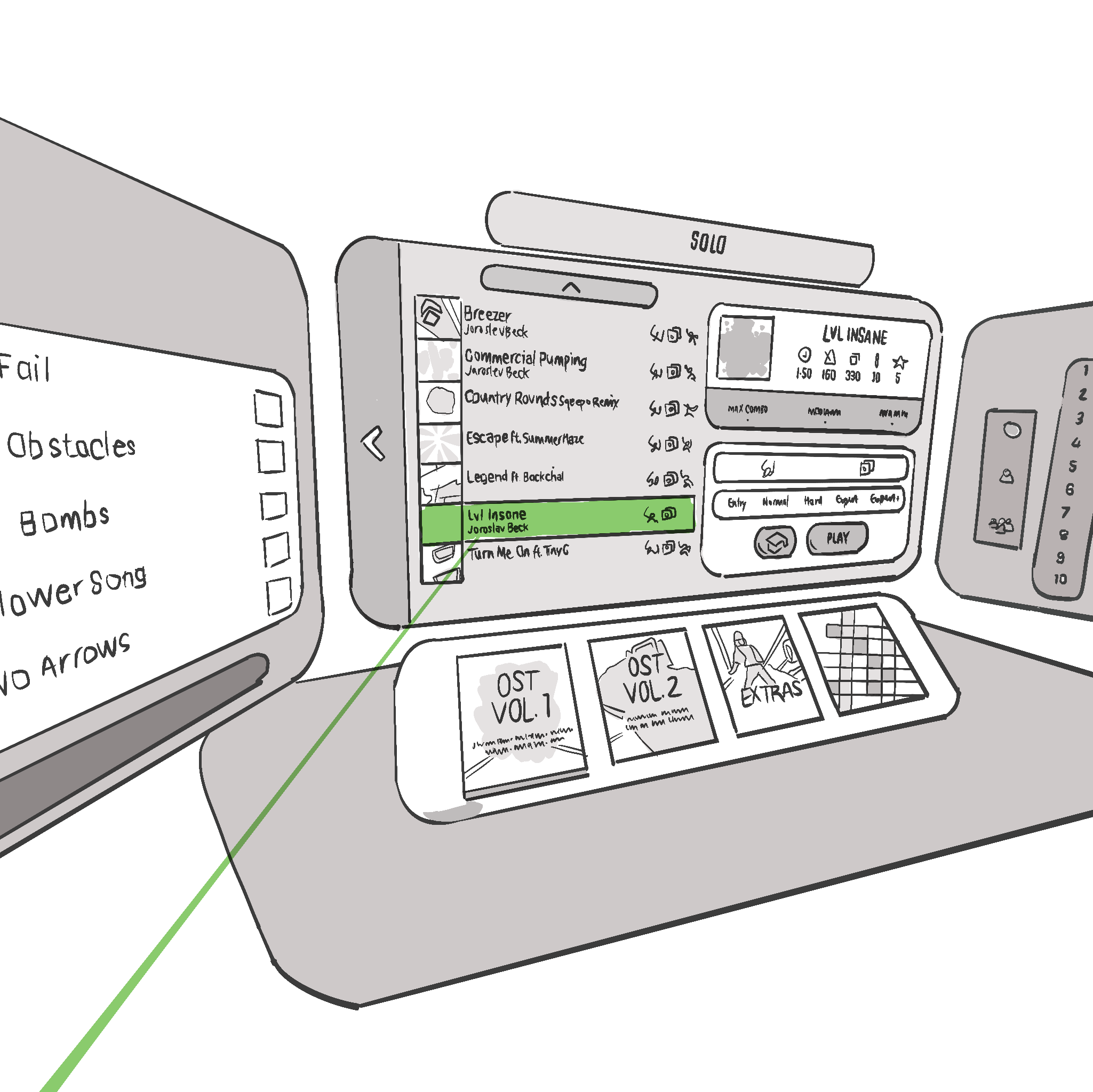}
    \caption{Multi-screen non-diegetic menu screen in \textit{Beat Saber}}
    \label{fig:menu1}
\end{figure}

3DUIs have excelled at emphasising direct manipulation of interface elements. Indeed the over-arching metaphor of most applications is that the environment affords opportunities to explore, experience situations and then control some aspects of the environment by actions such as manipulation. In many cases, an application might solely be focused on the user interacting with the environment. However, there is often a need to configure aspects of the experience or call up controls to transition to another location. Such features usually result in the developer constructing some sort of pop-up 2D interface that resembles a dialog screen that one might encounter on a traditional 2D window. These 2D interfaces can pack a lot of controls and are thus quite an efficient use of screen-space. They might include tabs, scrollable windows, sliders, buttons and many other recognisable features. Of course, these features might not be as easy to use in 3D. There are a vast array of options for rendering and composition that are beyond the scope of this article, ranging from flat menus that would not look out of place in a mobile app (e.g., \textit{\textbf{Beat Saber}}, see Figure \ref{fig:menu1}), to 2.5D menus that are mostly flat, but with 3D relief to look like physical controls such as buttons and sliders (e.g., \textit{\textbf{Rec Room}}) through to menus laid out on 3D surfaces that are more effective (e.g., the handheld menu in Google's \textit{\textbf{TiltBrush}}, see also Section~\ref{sec:inhand}). This latter example also shows the power of using two hands over one for menu interaction tasks~\cite{Lindeman1999a}.

\begin{figure}
    \centering
    \includegraphics[width=0.5\linewidth]{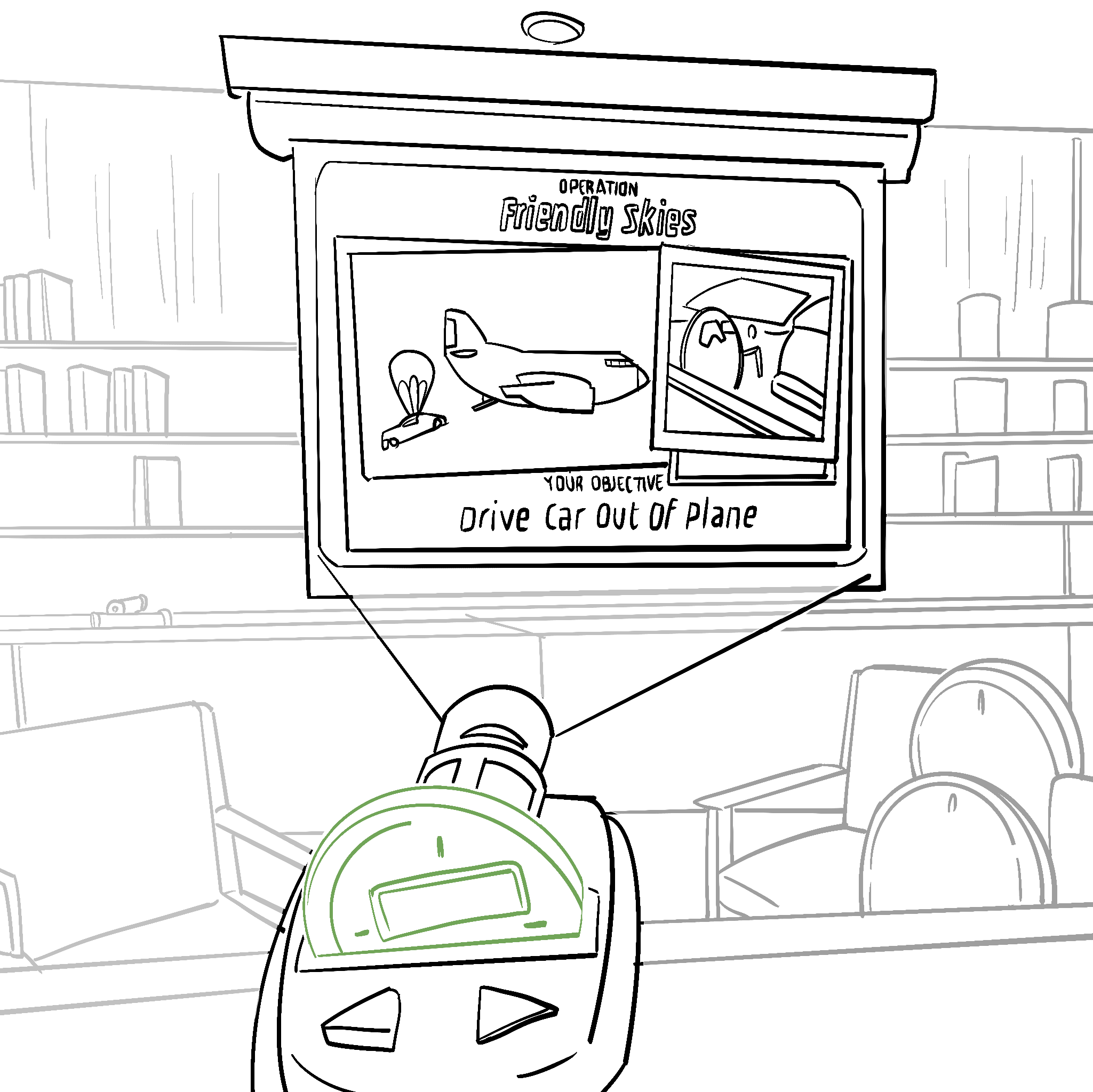}
    \caption{The projector as diegetic metaphor for a menu in \textit{I Expect You to Die}}
    \label{fig:menu2}
\end{figure}

The balance for how much to put into the environment or into a menu overlay is very much up to the designers. Some have pushed as much as possible into the environment, so that the user never need encounter overlays or pop-up windows, whereas others make liberal use of menus. This distinction is sometimes referred to as \emph{diegetic} versus \emph{non-diegetic}, with the use in consumer VR being very similar to the description of the use of the term in first-person shooter games~\cite{andrews_game_2010}. For example, the game \textit{\textbf{I Expect You To Die}} has multiple levels, so an obvious way to select levels would be to put up a menu where the user selects them. Instead, the game provides a diegetic metaphor where levels that are available appear as film canisters that can be placed into a projector, see Figure \ref{fig:menu2}. In a tongue-in-cheek metaphor, the game \textit{\textbf{Job Simulator}} presents its level selection mechanism as choosing between game-console cartridges in a tray, and having the player insert them into a console to load them. The mechanism to exit the level is to take two bites from the ``Exit Burrito,'' which stretches the diegetic metaphor, but avoids a pop-up menu. 

\subsection{Physics-based Interaction}

The issues of following a diegetic pattern or not is strongly related to the issue of whether the virtual environment should follow rules from the real world. This is not a new problem, and the related issues have shaped 3DUI design from the earliest days. The most obvious issue is that virtual objects cannot have physical properties, and thus cannot exert forces on the user. Virtual objects are thus not ``solid'' when reacting to touch from the user, be it a virtual tool affording the form of a grip, or a virtual chair supporting the weight of the user. A secondary issue is whether objects obey gravity and other kinematic behaviours, so that while they cannot react to the user completely authentically, they can react with each other.

What is perhaps framed differently for consumer interfaces, is that the repercussions of any lack of physical constraint generates a variety of reliability issues and edge cases. Physics simulations are by their nature difficult to predict and thus random effects might be expected. Typical issues include objects falling out of reach, objects disappearing off to infinity because they escape collision volumes, and glitches with collision (e.g., stuck objects). This makes creating a reliable simulation very difficult; whereas in a lab situation a simulation engineer (or student) might reset a problematic object or level, a consumer experience has to be self maintaining, or provide for the user to reset things when needed. 

We might argue that the solutions to this are generalisable issues, and thus some solutions (e.g., making useful objects that are occluded appear to shine through their occluder as if in x-ray vision, as happens in \textit{\textbf{I Expect You to Die}}) are underscored in later sections, but our feeling is that this is a more general issue of how to determine whether the application is actually in a state where it can progress, and thus is more a systems or programming issue. Indeed, it hints at an interesting theoretical issue of testing about whether a goal is achievable from the current state represented within the game engine. 

Finally, we note that consumer applications span the whole range from relying on simulated physical behaviour (e.g., \textit{\textbf{Gadgeteer}}) through to completely fantastic environments that exploit movement, but ignore physics (e.g., \textit{\textbf{Beat Saber}}).

\subsection{Likely changes}

As a final framing issue, we note that while consumer content has mostly evolved to fit with the dominant types of controller (two 6DOF controllers with a range of buttons), we might expect a variety of changes over the coming years. Generally, new modalities operate as a superset of previous modalities. Thus, while Google Cardboard-style interfaces supported gaze only, or gaze+button interaction, experiences using just these are still playable on later systems, subject to software compatibility. 

It seems reasonable that the current state might remain the ``entry level'' virtual reality experience for a few years. Features such as eye tracking and body tracking can enable new ways of interacting with objects, but they do not need to deprecate older ways of doing things. Hand tracking, though, has a slightly different use, as evidenced by the HoloLens devices, in that it can replace the use of a controller. However, this does not appear to fit with the needs of VR content at the moment. It would be difficult to map all the buttons and joystick input of a controller to hand gestures. Further, a hand controller easily enables multiple actions at the same time (e.g., moving, pointing and grabbing), while the hand gestures to enable different modes might only operate exclusively. Timing and precision of hand gestures is also a poor replacement for the accuracy of a controller, due to the physical nature of the device. Thus, hand tracking raises a range of new issues about the operation of emerging types of consumer content.

\section{Review Methods \& Taxonomy}
\label{sec:taxonomy}

There are already thousands of consumer applications and demonstrations across the various VR platforms. While some of the main platforms have heavily curated content that focuses on high quality, there are very active independent developers creating either for the more open platforms (e.g., Steam/Windows, or Android), or focusing on hobbyist markets (e.g., Oculus Quest via SideQuest). Each experience might offer a variety of different user interface techniques, possibly selected by the user. Some experiences may not reveal novel techniques until a substantial amount of experience has been gained, or might only be enabled in certain device configurations. A user might not even notice that a particular interface technique is not something ``standard.''

Even in relatively narrow domains such as manipulation, it is practically impossible to systematically review all the options, and thus claim to be able to categorise and analyse them all. As we have shown in Section \ref{sec:framing}, interaction techniques are also dealing with new issues and thus it remains unclear if older taxonomies can capture the variety of issues to which developers and designers today are responding. 

Thus, in doing the background research for this paper, the key idea by the initial team (Steed, Lindeman and Johnsen), was to crowd-source examples of interaction techniques. A Slack channel was created as a working space, along with a Trello board to record ideas. Various social media routes were taken to advertise the crowd-sourcing effort, including advertising to community email lists (3DUI list, IEEEVR list), a number of private Slack  channels (including one for VR game developers and another for distributed VR experiments) and social media posts on Twitter. The process was open, with the Trello board serving as a discussion board initially, then a todo list for the writing of this paper, and now as a living resource for readers to engage with.

We had discussions about whether to follow older taxonomies of 3DUIs, what comprised a useful transferable result and whether specific techniques were transferable or not (c.f. discussion of \textit{Exit Burrito} in Section \ref{sec:framing}). A particular point was whether we should spend a lot of time looking for prior academic art for every technique. In the end, the decision was made that because this was a crowd-sourcing project, we should not try to make a scientific argument about the benefit of a specific technique, and thus while prior art might be found, it would not invalidate the raising of the example as a design exemplar. Thus, while we have made links to related research when appropriate, we have not attempted to systematically compare to prior art. We do encourage readers to add references to related work on Trello (see Section \ref{sec:conclusion}).


The Trello board contains many links to video fragments that illustrate the key points in the following sections. Examples that were added after this paper will be clearly labelled. The discussion on Slack is still open and we welcome contributions (see Section \ref{sec:conclusion}). 

\section{Selection}
\label{sec:selection}

In this section we discuss novel implementations or insights into the task of selecting objects. This is a short section, as there are relatively few ways to implement selection in an immersive system, and it is usually combined with techniques for manipulation that are discussed in the next section. It is more important for complex environments where the intentions of the user are hard to interpret, as selection is needed to match a somewhat complex gesture of the user (e.g., pointing) against a scene where it can be difficult to match the target object unless there is some understanding of the role of the user (e.g., see \cite{steed_towards_2006}). The ``standard'' techniques of reaching and touching, or pointing with a ray have seen a lot of adaptation to specific situations. The concept of a bendy ray was already quite well explored in the 3DUI community~\cite{kai_riege_bent_2006} and is commonly used in consumer applications. Thus, we consider it out of scope for further discussion and in the spirit of this paper, we describe three examples of techniques suggested from our crowd-sourcing exercise that highlight opportunities for further development.

\subsection{Intention-preservation Selection}

Selection can be broken down into two separate low-level tasks: Indication and Confirmation (analogous to Pointing and Clicking). Indication is typically done using ray casting or direct touching, while confirmation is often done by pulling a trigger, squeezing the grip or making a fist. When indicating an object out of arm's reach, a simple technique is to project a selection object (such as a ray, cylinder or capsule) from the user's hand out into the environment. If this selection object is visible, then indication is relatively straightforward for the user, as there is direct feedback about which object will be selected. However, it might not be so obvious in which direction the selection object will point depending on the hand shape. This has led to other techniques such as eye-hand ray selection, where the object behind the hand from the user's view is selected. Questions about which might be preferred have been debated in several papers. We refer the reader to Argelaguet \& Andujar's paper describing the drawbacks of both~\cite{argelaguet_efficient_2009}.

In creating the game \textit{\textbf{Bullet Train}} several decisions about how to implement object selection and grabbing had to be made. Donaldson \& Whiting discussed this in a talk at the Virtual Reality Developers Conference in 2016~\cite{donaldson_going_2016}. The game is a fast-paced shooter where the user picks up various weapons. In order to prevent too much bending over, weapons are picked up at short distances ($<2m$). Effectively, they are selected first by collision with a capsule attached to the hand, where moving the hand is used to indicate which item will be selected, and then can be grabbed using the trigger. In testing, they found that neither hand-pointing nor hand-eye pointing preserved the intention of all users, in that different users would try to select the items in different ways. Their suggestion was to use the vector between hand-eye and hand-pointing, but to switch to hand-pointing when the user pointed ``from the side.'' The explicit metric used to switch was not given in the talk, but it would make sense to have a threshold on both the angle of the hand from the eye direction, and the forward direction of the hand from the eye direction. 

\begin{figure}[h]
    \centering
    \includegraphics[width=0.6\linewidth]{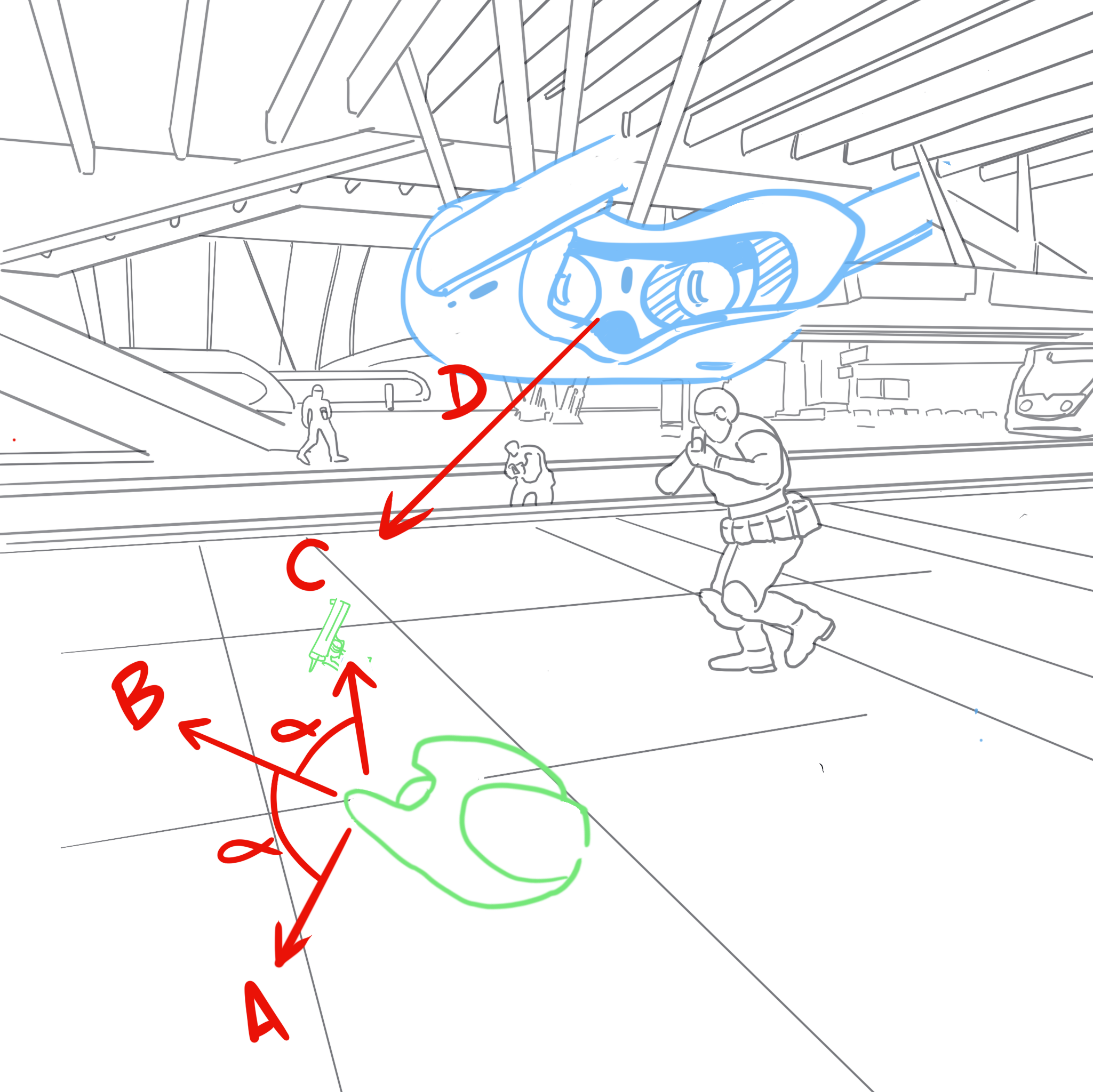}
    \caption{Selection ray in the \textit{Bullet Train} game. Vector $A$ indicates the eye-hand vector. Vector $C$ indicates the hand forward direction. Vector $B$ indicates the compromise direction that preserves the intention in most cases. Note: In this and future illustrations, blue elements are superimposed representations of the interface devices themselves. Red and solid-black lines are annotations. Green items are the in-environment representations of the controllers.}
    \label{fig:selectbullet}
\end{figure}

This is illustrated in Figure \ref{fig:selectbullet}, where Vector $B$ bisects the forward vector of the hands (vector $C$ in the figure) and the eye-hand vector (vector $A$). The criteria to revert to hand pointing (Vector $C$) would be if Vector $D$ (the head direction) and Vector $C$ differ by a threshold. Note the exact criteria is not discussed in \cite{donaldson_going_2016}, and might just be an angle threshold. 

We note the similarity of this technique to other explorations of how to infer intention with the hand depending on the geometry of the scene. For example, Wagner et al. explore combining hand and ray selection depending on scene structure~\cite{wagner_comparing_2021}).  The generalisable pattern that might be explored is how the intention of the user can be preserved given knowledge of the environment and the selection actions being performed. A domain where this has been explored before is sketching systems and modelling for VR (e.g., \cite{jackson_lift-off_2016, baerentzen_signifier-based_2019}), where the action space is very large, but can be significantly constrained by the local context. What \textit{\textbf{Bullet Train}} suggests is that, even in simple applications or for simple actions, intention might be a more exploitable notion.


\subsection{Combined Head and Hand Selection}

\begin{figure}[h]
    \centering
    \includegraphics[width=0.5\linewidth]{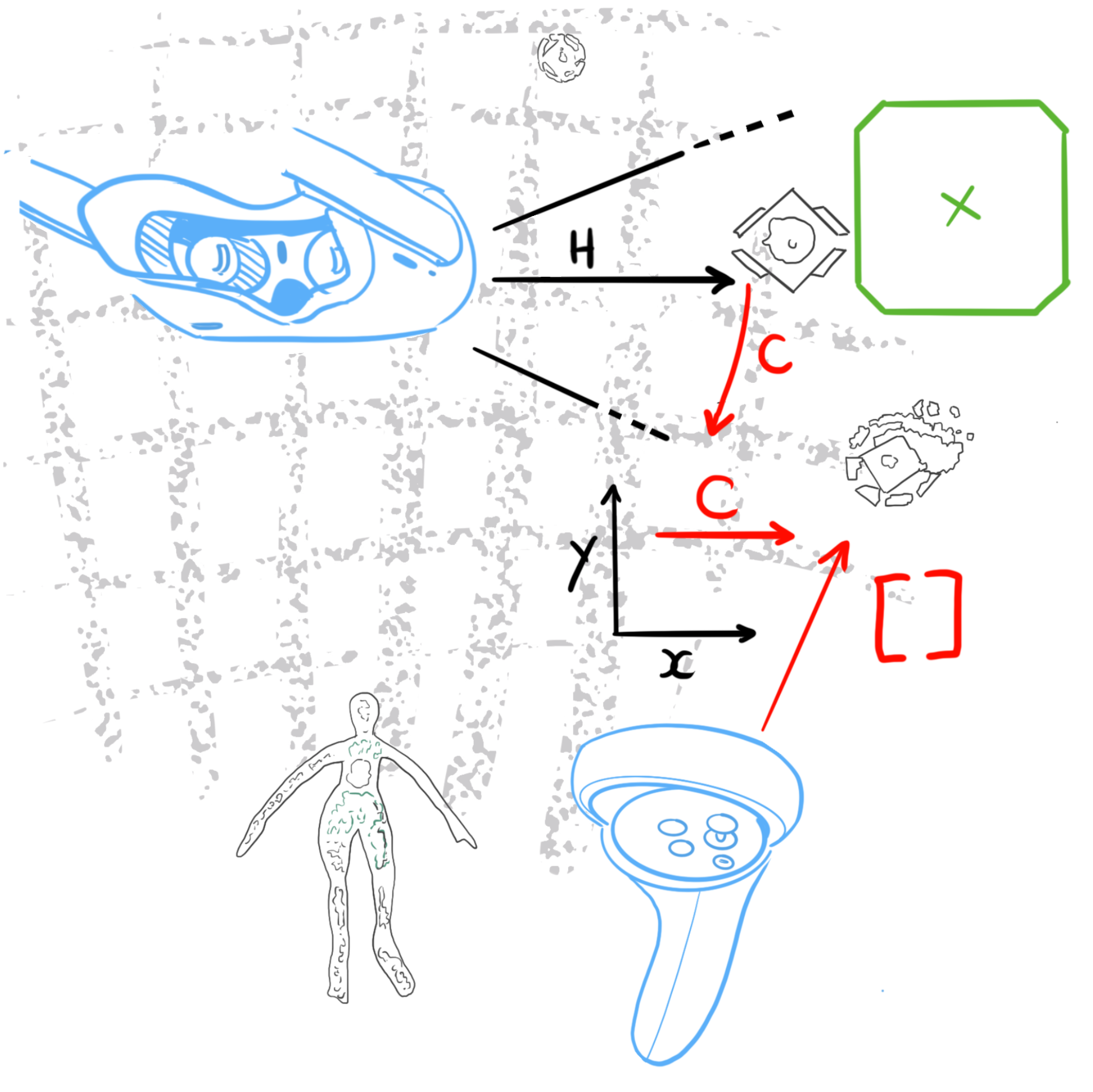}
    \caption{Combined head and hand selection. The selection crosshair position (Green) is defined by the head direction $H$ and modified by a vector $C$ on the image plane. This vector can be built from either a thumbstick or using Yaw and Pitch angles from a tracked controller. It is also clamped to stay within the headset FOV. }
    \label{fig:selectrez}
\end{figure}

Another way of improving the usability of selection techniques in VR is to provide multiple ways to control them.  In the game \textit{\textbf{Rez Infinite}}, the developers provide the option for a selection method that combines head movement and controller input. 
The core game mechanic consists of selecting and destroying targets while the camera moves at constant speed through the levels. 
In the PlayStation VR version of the game, the controller input is combined with head orientation: the crosshair can be controlled by turning the head or using the thumbstick on the controller, see Figure~\ref{fig:selectrez}. The game also works with the PlayStation Move controllers, or other 6DOF controllers on other platforms, by using the yaw and pitch angles. Combining hand and head input has a series of advantages. Besides allowing users to choose the method they are most comfortable with, they can also use both of them simultaneously or in sequence. Head pointing makes it easier to select a single object on screen, while the controller allows the user to make quick multi-selections when objects are lined up. It is also possible to use the head to drag the cursor closer to the object of interest and then refine the pointing by using controller input. There are also ergonomic considerations; although pointing with the head is more intuitive, it can be physically demanding. The controller, on the other hand, can stay in a comfortable position during the game, since only yaw and pitch rotations are used to control the crosshair. The technique used by \textit{\textbf{Rez Infinite}} has some limitations. Since the head  controls the cursor, it is not possible to look at other objects of interest without drifting away from the current selection group. In addition, since head and controller move independently, the cursor cannot keep a fixed relative position to either one. Sometimes pointing the controller forward will select objects in the centre of the FOV, and sometimes not. This limits the advantage of the natural pointing capability of tracked controllers.

The general consideration here is that in many more abstract applications, selection is the primary goal and thus there is no need to delegate it just to gaze or hand, but to use both if it is efficient. This technique is highly amenable to testbed-style performance evaluation. 

\subsection{Gaze-dwelling Selection}

\begin{figure}[h]
    \centering
    \includegraphics[width=0.5\linewidth]{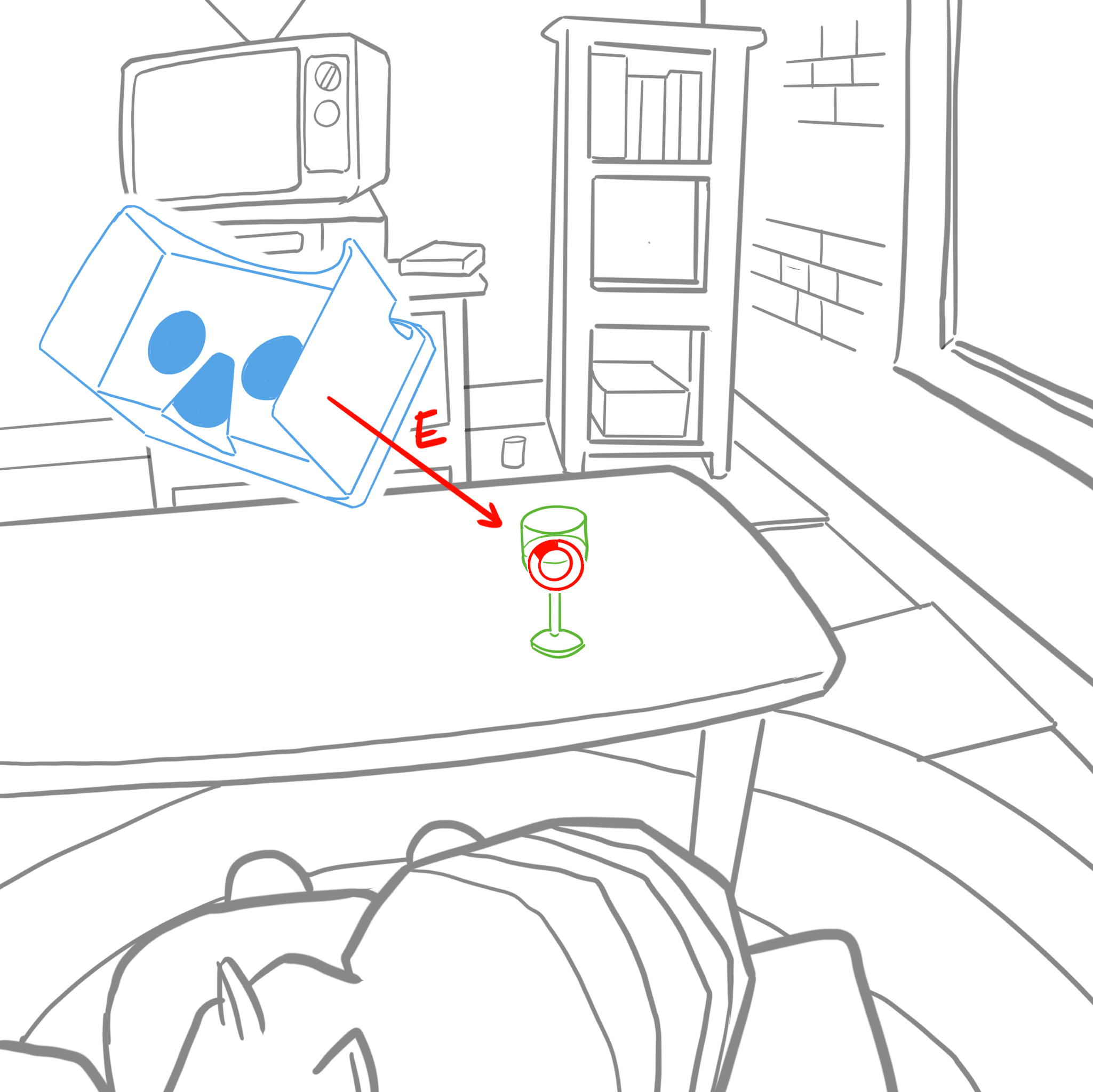}
    \caption{Gaze-dwelling selection. Head orientation is used to determine a vector for indication ($E$). The object is selected after the dwelling time has transpired. A circular bar or other visual method is used to indicate how much time is left before the object is selected.}
    \label{fig:selectdwell}
\end{figure}

In some games, in particular those designed for smartphone-based VR, selection is often accomplished by using a gaze-dwelling technique. Gaze dwelling is a well-known technique that uses the head orientation or eye gaze to define a pointing direction \cite{isokoski2009gaze,sibert2000evaluation,tanriverdi2000interacting,velloso2016emergence}. Objects are then indicated using ray-casting or another intersection test. To avoid the ``Midas Touch'' problem \cite{jacob1990you}, targets are only confirmed after a small fixation time, which is used to determine intent. The elapsed time is often communicated to the user through a circular icon that fills or another visual representation, see Figure~\ref{fig:selectdwell}. One example is the game \textit{\textbf{Wendy}}. In this VR witchcrafting puzzle game, gaze dwelling is used to select and interact with items in a room. Gaze selection can also be combined with travel. In \textit{\textbf{Manifest 99}}, gaze is used to pick up objects, and to select creatures and characters for the player to inhabit. In this game, character selection is indicated by a white shader that progressively fills the surface of the object being selected. Besides being a simple and intuitive technique, gaze dwelling creates a distinct experience from using buttons. The game \textit{\textbf{Land's End}} was initially released for Samsung Gear VR which lacked controllers. However, when it was released for Oculus Go (which has a 3DOF controller), the designers decided to stay with gaze dwelling to maintain the experience of moving and solving puzzles only with the eyes \cite{pashley_revisiting_2018}. 

The observation here is that eye-gaze dwell is indicative of user interest and could be a viable selection mechanism in many situations. It also has potential use in future accessible interfaces. The main limitation of this technique is the time required to complete the selection, which limits its use to slow or infrequent selections \cite{chen2019using}.

\section{Manipulation}
\label{sec:manipulation}

Somewhat similarly to selection, there are not so many options for manipulation with the common configuration of two hand controllers. The obvious approach is to attach the object to the hand~\cite{mine_moving_1997}. However, as we discuss in the following three sub-sections, there are some interesting new variants that have emerged that deserve further study and use.

\subsection{Tomato Presence}

\begin{figure}[h]
    \centering
    \includegraphics[width=0.3\linewidth]{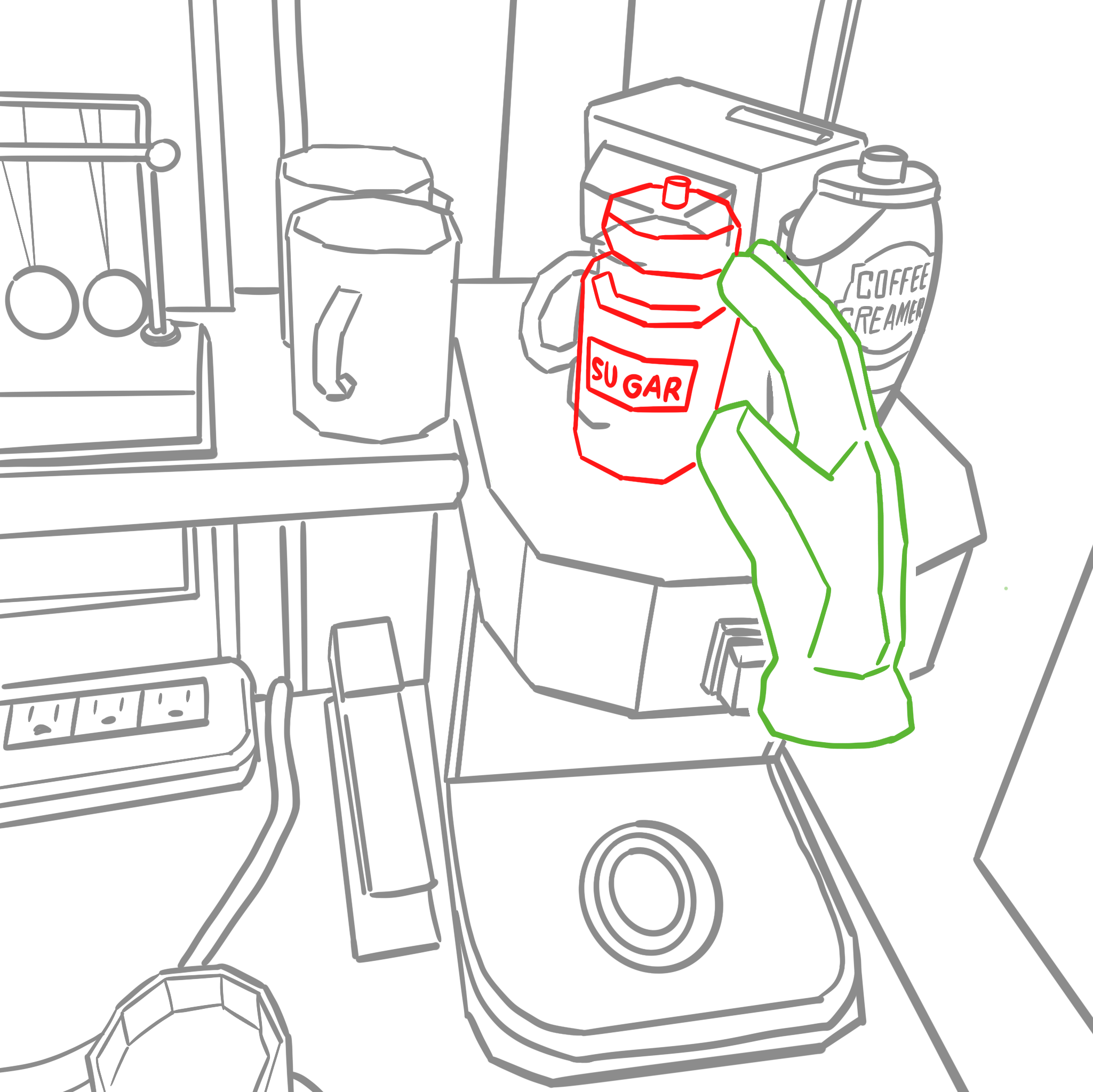}
    \includegraphics[width=0.3\linewidth]{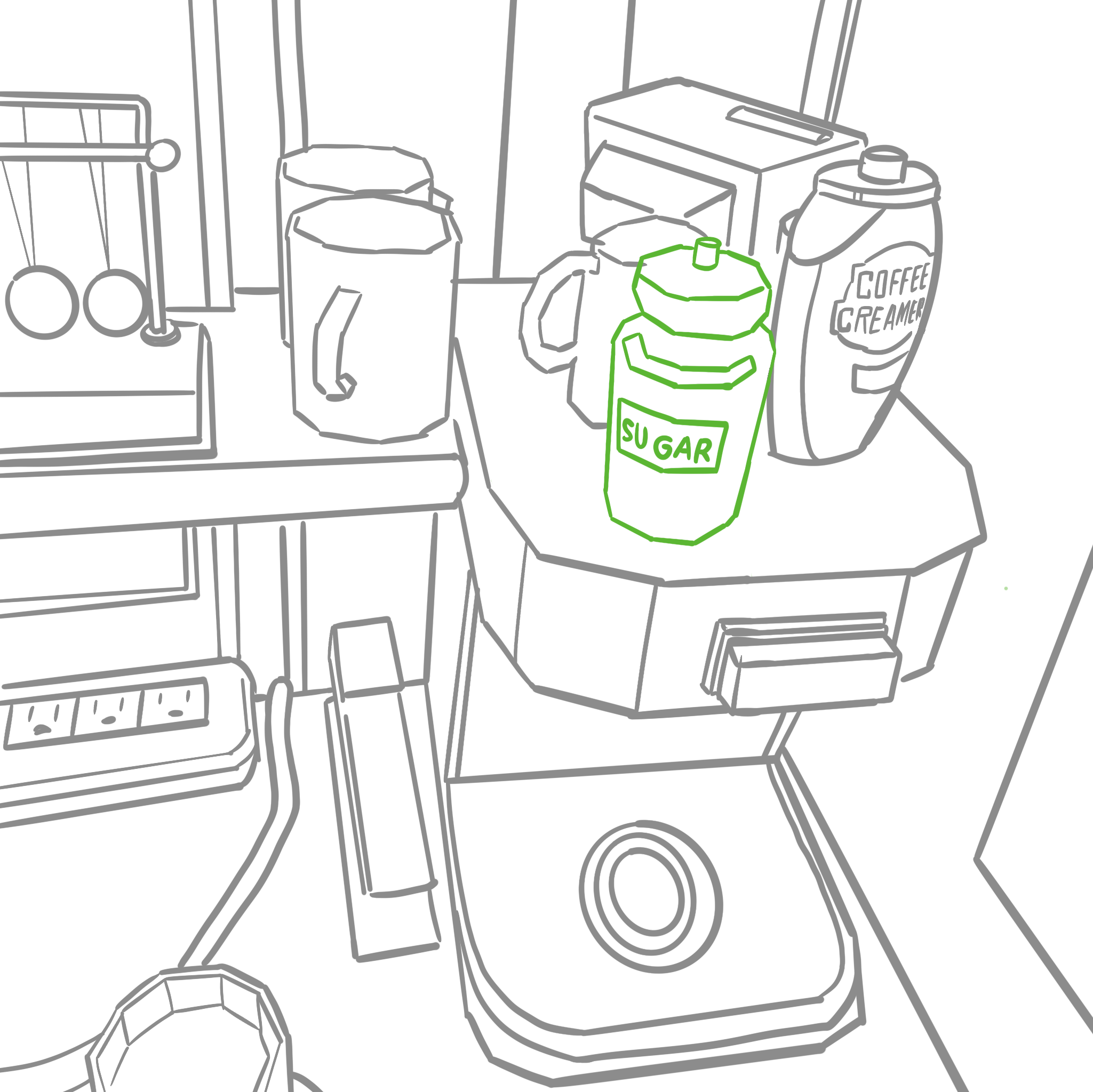}
    \includegraphics[width=0.3\linewidth]{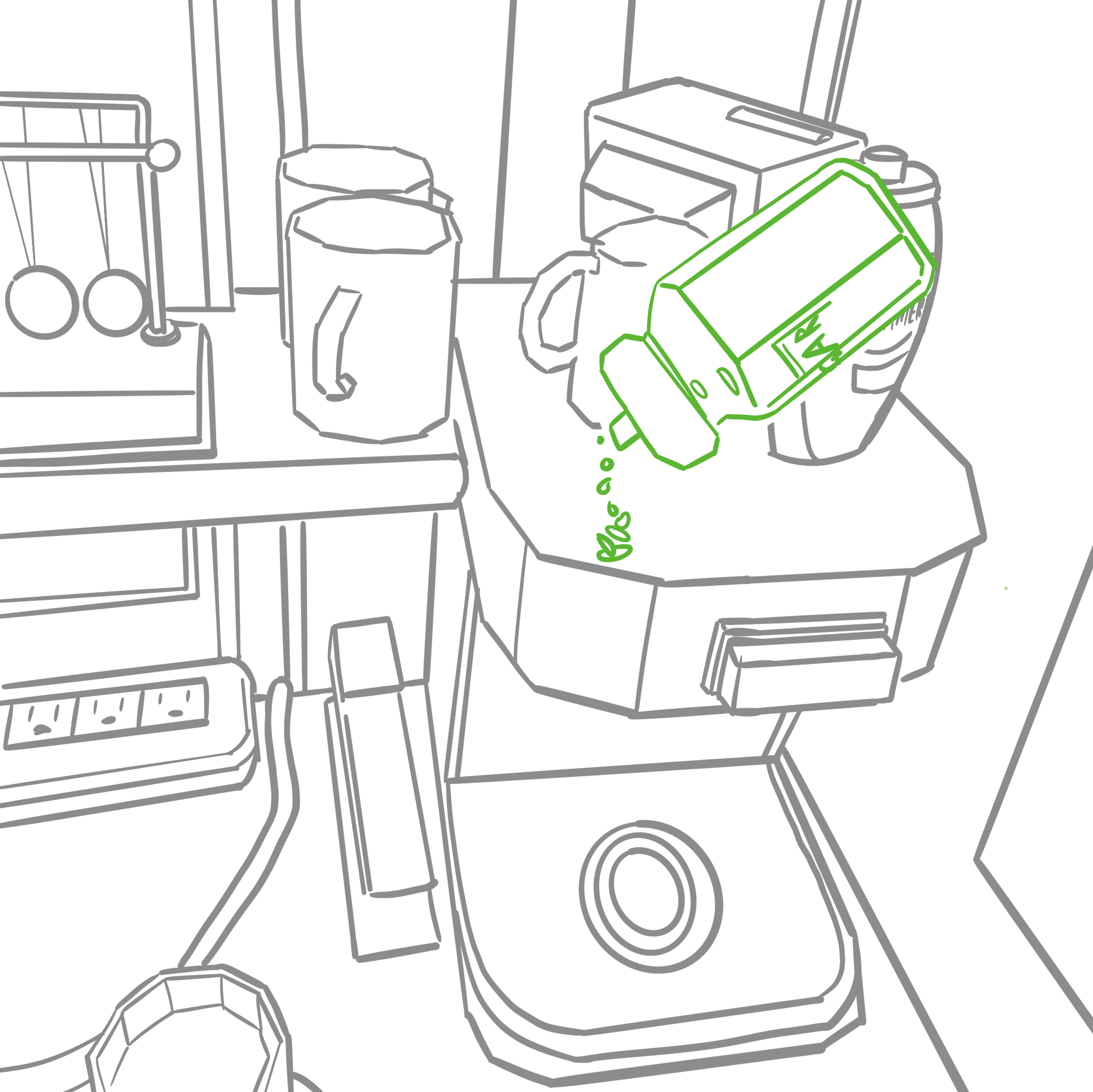}
  
    \caption{Tomato Presence in action. As the user touches the sugar shaker (left), it changes colour. As they grab it (middle), their hand disappears. They can thus manipulate the object (right). Releasing the object reverts back to their original hand.}
    \label{fig:tomato}
\end{figure}

The term \emph{tomato presence} was coined by Alex Schwartz of Owlchemy Labs \cite{owlchemy_labs_tomato_2020} as a description of the grabbing mechanism in their game \textit{\textbf{Job Simulator}}. The concept is very simple: when the user grabs an object, their hand disappears, and becomes the object they are manipulating (see Figure \ref{fig:tomato}).

This game involves a lot of object manipulation in fairly close quarters. Tomato presence avoids a common problem, which is how to show the hand of the user grabbing an object; either the object is shown somehow just stuck to the hand, or the object has to be moved into a position where the hand model can be shown to grasp the object. The latter is commonly done in games, but it does mean that the grasp shapes have to be created for many objects and grab configurations.

The technique might be considered controversial because of the prior work that shows that users feel embodiment in avatars with human-like hands~\cite{spanlang_how_2014} and that embodiment can be lower when the user sees an arrow not a hand~\cite{yuan_is_2010}. However, the consequences of moving the hand are still visible as the object moves. Owlchemy Labs have claimed that few users notice that their hand disappears\cite{owlchemy_labs_tomato_2020}. 

Tomato presence is potentially very widely applicable in the most common scenario where the user is represented only by their hands. It would obviously be more difficult to justify if the whole body was being represented. We would highlight the connection to recent work about how avatars can help or hinder interaction. For example, some work has shown that making the avatar partly transparent might aid in typing \cite{knierim_physical_2018} or make more ``avatar friendly'' interaction techniques \cite{dewez_towards_2021}.

\subsection{Snap Drop Zone}

One problem that can arise when manipulating objects is what to do when they are released. Two obvious solutions are just to let them fall to the ground (i.e., obey physics) or to leave them floating in the air (i.e., ignore physics). While some  might claim that the ``right'' choice would be the former, others would claim that the latter makes for better game play, especially for situations where the player needs to reacquire a released object quickly.

\begin{figure}[h]
    \centering
    \includegraphics[width=0.5\linewidth]{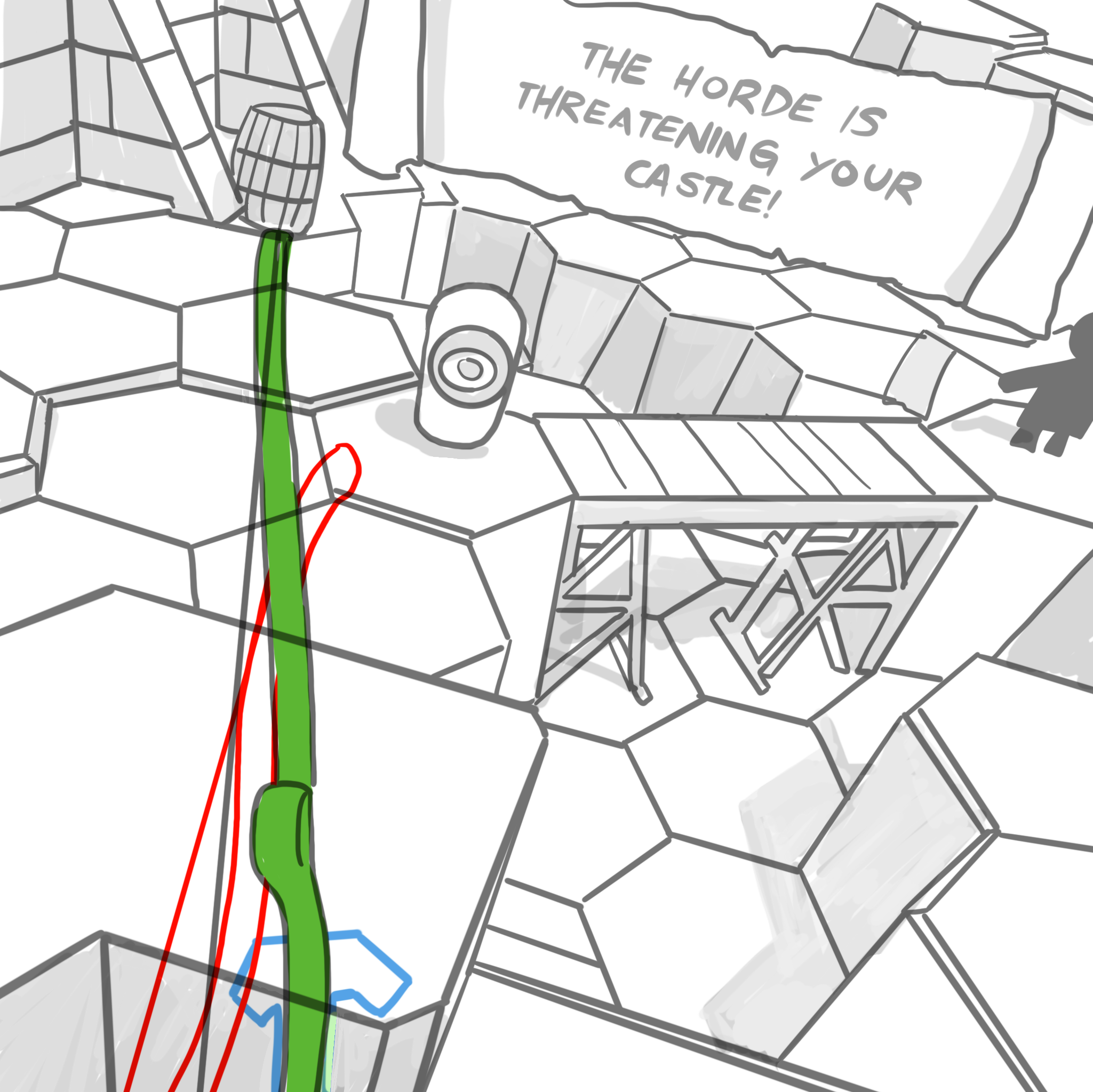}
    \caption{Snap Drop Zone in the HTC Vive \textit{Longbow Tower Defence} game that comes as part of \textit{\textbf{The Lab}}. The outline of the bow is highlighted when the controller touches it, and releasing the grip button snaps the bow back into place against the wall.}
    \label{fig:select}
\end{figure}

A middle ground approach is to snap the object back into the location and orientation where it was first picked up. This Snap Drop Zone technique (Figure~\ref{fig:select}) can be very useful, especially for situations where the user stays stationary for some time or is in a vehicle. But whatever approach is chosen, the main point is to build \textit{predictability} into the experience, so that the behaviour doesn't break the sense of enjoyment or engagement.

While snapping techniques are somewhat common, especially in construction or engineering applications (e.g., see \cite{mapes_two-handed_1995}, Snap Drop Zone is potentially widely reusable because the object can always be found in one or a small number of places if it isn't carried. This makes it useful for tool-based interfaces. We can also envision this technique being good for things that the user may carry around. For example, if there is a virtual mobile phone, tablet or weapon that might be holstered. Other options are possible, but again, predictability is key.

\subsection{Reachability}

As discussed in Section \ref{sec:dealing}, a particular problem for the consumer VR market is that the application developer has to deal with different sized spaces being available. Many early games were designed to be played from a seated or stationary position because the systems did not have head tracking, or only had limited-range head tracking (e.g., Oculus Rift DK2). For example, in the game \textit{\textbf{I Expect You to Die}}, the scenes are created for stationary experiences, often from a virtual seat. Thus, reachability is not so much an issue, though interestingly that game does have a mechanic for picking up objects that fall out of sight, due to the natural physics used; objects that are needed but are out of sight are highlighted through the object they fall behind. In contrast, some games require larger spaces. For example, the game \textit{\textbf{Unseen Diplomacy}} makes the maximum use of a minimum 4m x 3m tracking space. 

For more general environments, bringing everything into reach from a single position is perhaps not possible, but bringing everything within the chaperone or guardian boundary might be. \textit{\textbf{Job Simulator}} has three different layouts for each of the game scenes depending on the space that is available as reported by the tracking system and chaperone boundary \cite{owlchemy_labs_room_2016}. Thus, the game will work in a space as small as 2m x 1.5m. This required careful thought by the design team to structure and position key interaction work spaces (e.g., counter tops, dish washers) in a relative fashion.

The general observation is that prior work has focused on usability without consideration of reach and interaction with a boundary. Beyond simply using the chaperone, the physical environment could be modelled by scanning~\cite{sra_procedurally_2016}. During the writing of this paper, the chaperone system of the Oculus Quest gained an experimental feature to mark up couches~\cite{heaney_you_2021}. Thus, we should expect more mapping information to be available to the application developer in future toolkits, and it will be interesting to see how this can be exploited by researchers and developers. There are already interesting demonstrations of the possibilities here (e.g., rendering certain real objects into the scene~\cite{hartmann_realitycheck_2019} or procedurally generating environments to match tracking spaces~\cite{sra_procedurally_2016}).

\section{Locomotion}
\label{sec:locomotion}

Probably one of the most widely-studied interaction types within the VR academic community is locomotion \cite{steed_displays_2013, boletsis_new_2017, di_luca_locomotion_2021}. A locomotion technique attempts to address the problem of mapping movements within a finite physical space to player movement within a potentially infinite virtual space. One way to tackle locomotion is to build a device on which the user walks. This has been the topic of several academic efforts, including robotic tiles~\cite{circulafloor}, and omnidirectional treadmills~\cite{torustreadmill}. Such devices are unlikely to be consumer systems in the short-medium term, but there has been a variety of devices for constrained walking, where the user takes physical strides, but either holds themselves in place or is held in place by a harness (e.g., \textit{\textbf{ROVR}}~\cite{swapp_implementation_2010} or \textit{\textbf{Cyberith Virtualizer}}).

For the vast majority of users and demonstrations, the developer will need to provide a locomotion technique based on controller inputs and motion tracking. In this section, we report eight of the techniques reported by the crowd-sourcing effort. We refer the reader to the Trello board (see Section \ref{sec:conclusion}) for similar discussions of a further nine (at the time of writing) similar techniques.


\subsection{Point-and-Click Walking}

The first-person game \textit{\textbf{VRZ: Torment}} contains several locomotion options, one of which is a hybrid between teleport and steering locomotion. The player chooses a destination location with a curved selection arc as is typical in teleport locomotion, but instead of teleporting to that destination, the player viewpoint starts automatically walking towards the destination after the selection is complete. This comes with the downside of the automatic walking being virtual movement, which increases the risk of visually-induced motion sickness. The upside of this form of locomotion is that it is more realistic than teleporting and also requires little effort while the viewpoint is moving towards the destination. 

Teleport is now a very common option in consumer VR applications, but there is a worry that it leads to a lower spatial awareness as the user does not have the visual experience of travelling \cite{macquarrie_effect_2018,rahimi_scene_2020}. Thus, the general observation is that techniques such as point-and-click walking might prove to be a happy medium, and might also lead to better spatial understanding by the player than teleportation, because users get a visual transition and can look around during the automatic walking process.

\subsection{Narrowing the Field of View}
\label{sec:narrow}

\begin{figure}[h]
    \centering
    \includegraphics[width=0.5\linewidth]{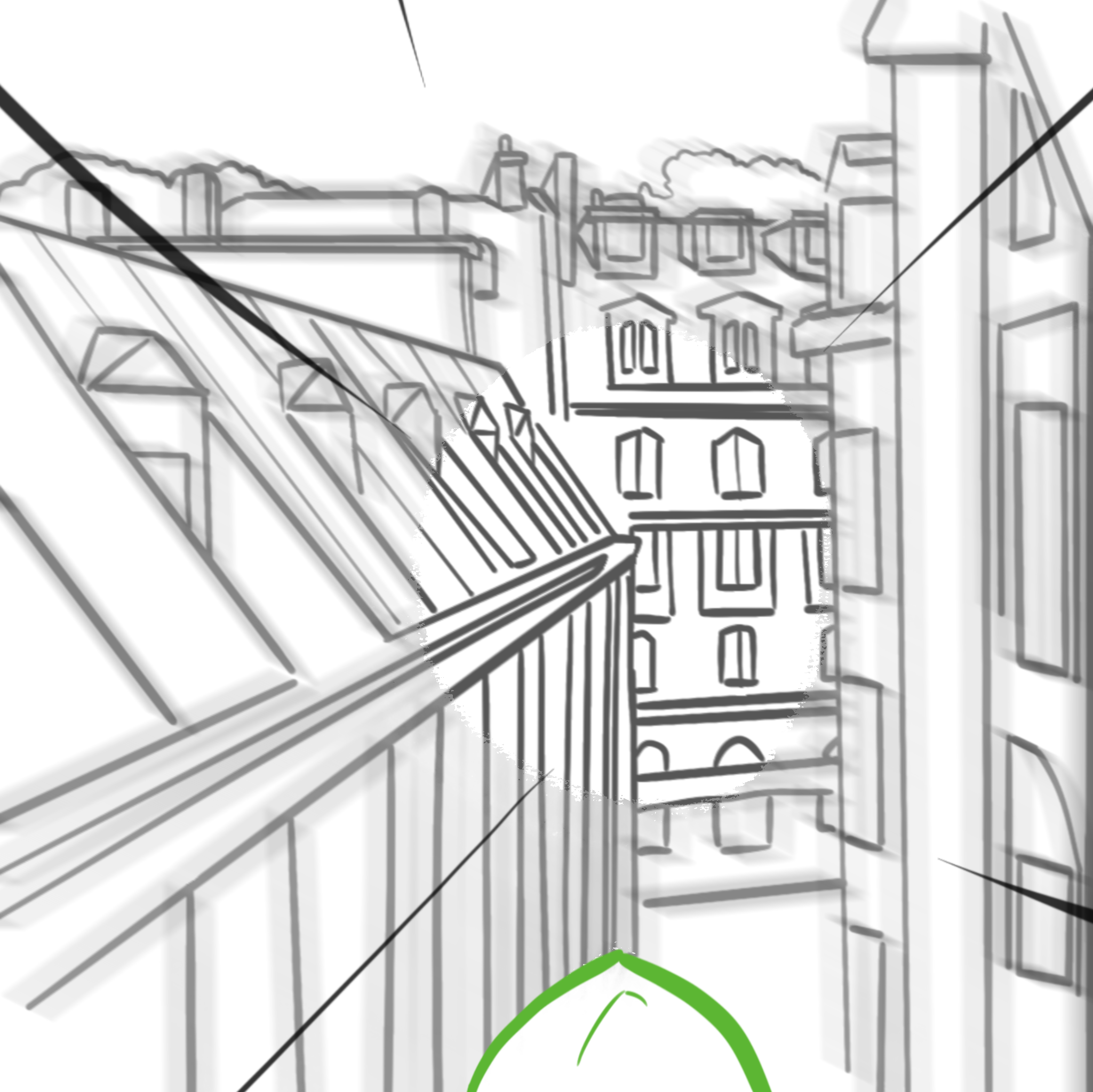}
    \caption{Narrowing field of view, as shown in \textit{Eagle Flight}}
    \label{fig:fov}
\end{figure}

The game \textbf{\textit{Eagle Flight}} is notable for violating an assumption that many have taken on board: that users would get nauseous if moving quickly and turning relative to the horizon. The game has the player flying as an eagle, with visual parallax quite close to the head and with the user performing steep rolling turns. When turning though, the field of view narrows by blurring and a vignette effect around the screen, see Figure~\ref{fig:fov}.

This appears to be a technique that could be applicable in environments with fast or turning motions. Certainly it seems to be an effect that might be configurable by the user depending on their sensitivity to motion cues. Some academic work and descriptions have recently emerged~\cite{bolas_dynamic_2014,fernandes_combating_2016,adhanom_effect_2020}. How broadly this can be used is still an open question.

\subsection{Blurred Reorientation}

Another technique, blurred reorientation, is similar to field-of-view restriction; optic flow is reduced via blurring, which subsequently could also reduce illusions of self-motion and motion sickness. In the \textit{\textbf{Cosmic Wandering VR}} demo, the whole screen is blurred during virtual view rotation, while in \textbf{\textit{VR Tunnelling Pro}} and \textit{\textbf{GingerVR}}~\cite{ang2020gingervr} developer tools offer functionality for blurring only the peripheral vision area.

Display blurring during virtual motion is a technique that can be used to reduce visually-induced motion sickness \cite{nie2019analysis}. One can imagine that if such techniques are proven to be broadly useful, they could become the responsibility of the interface software or hardware (similar in nature to the existence of chaperone systems) in order to facilitate accessibility, rather than being a feature that needs to be handled by individual developers. 

\subsection{Interaction Scale Adjustment}
\label{sec:scale}

Changing the scale of the user or the virtual world was explored in academic literature long before the current consumer VR boom \cite{kopper2006design}. Scaling the user into a virtual giant enables, for example, traversing large virtual worlds via natural locomotion, even in small physical play areas. Conversely, scaling the user into Lilliputian size gives an interesting new perspective and opens the door for more fine-grained object manipulation in spatial dimensions.

There are many ways to effect a player-scale change. For example, change in scale could be trigged by pressing a button on a hand-held controller, or automatically when the user enters predefined areas \cite{kopper2006design}. Krekhov et al. examined a locomotion technique where the user could switch between normal and giant mode to affect their stride length in the virtual world \cite{krekhov2018gullivr}. Their user study indicated that in comparison to standard teleport locomotion, their scaling-based locomotion technique enjoyed an increased presence and made the users walk around more frequently without causing cybersickness.

\begin{figure}[h]
    \centering
    \includegraphics[width=0.4\linewidth]{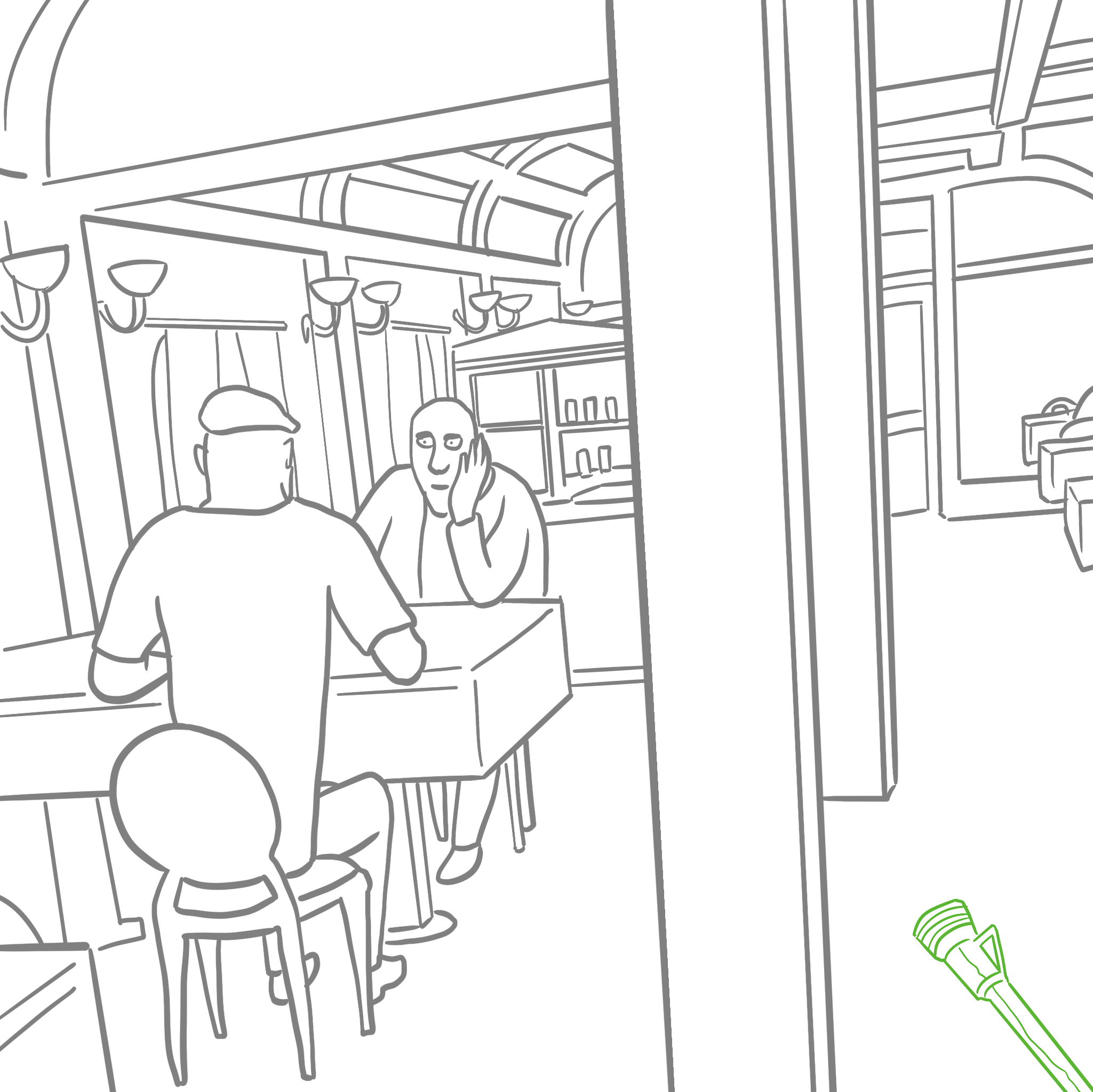}
     \includegraphics[width=0.4\linewidth]{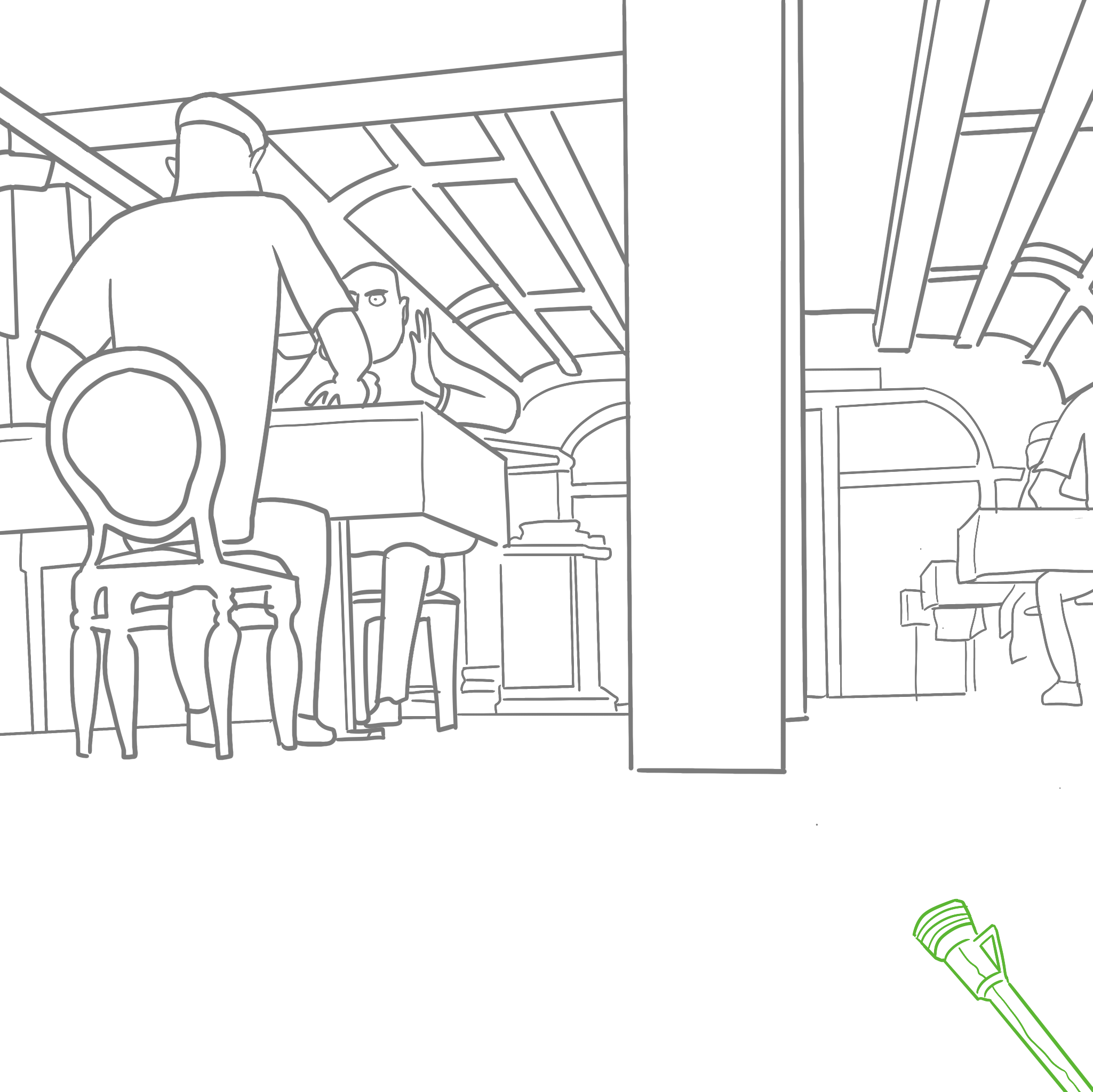}
    \caption{Scale adjust before and after from \textit{The Spy Who Shrunk Me}, via the shrink-ray tool.}
    \label{fig:scale}
\end{figure}

Examples of consumer VR applications with interaction scale adjustment include \textit{\textbf{Anyland}}, \textit{\textbf{NeosVR}}, \textit{ \textbf{Quill}} and \textit{\textbf{The Spy Who Shrunk Me VR}}. \textit{\textbf{Anyland}} and \textit{\textbf{NeosVR}} are social VR applications, where individual users can change their size and keep on interacting with others despite size differences. \textit{\textbf{Quill}} is a 3D painting application that allows the user to work at different scales. \textit{\textbf{The Spy Who Shrunk Me VR}} is a stealth game, where the player can shrink themselves to solve puzzles and avoid enemies (Figure \ref{fig:scale}).

We would make the point that scale changes, while interesting, might have different effects on user experience, such as changes in perception of distance, or even feelings of vertigo. Thus, this needs to be used judiciously, but in many environments there is a need to work on fine detail, as well as to get an overview. See also Sections~\ref{sec:wim} \& \ref{sec:space}.

\subsection{Rectangular Movement Gain}

This locomotion technique was developed for situations in which a virtual object is showcased in the centre of a virtual environment that is larger than the physical play area. It originated in the design for the \textit{\textbf{Audi AG Virtual Reality Experience}}. The idea is to selectively apply movement gain to allow the user to circle around the object of interest and look at it from different angles, even in cases where the object would not properly fit into the physical play area.

The technique works by enclosing the user within the bounds of a rectangle with a predefined size that is smaller than the physical play area and parallel with the horizontal plane. Whenever the user steps beyond this rectangle, movement gain is applied so that the virtual area ``slides'' in relation to the physical play area.

\begin{figure}[h]
    \centering
    \includegraphics[width=0.5\linewidth]{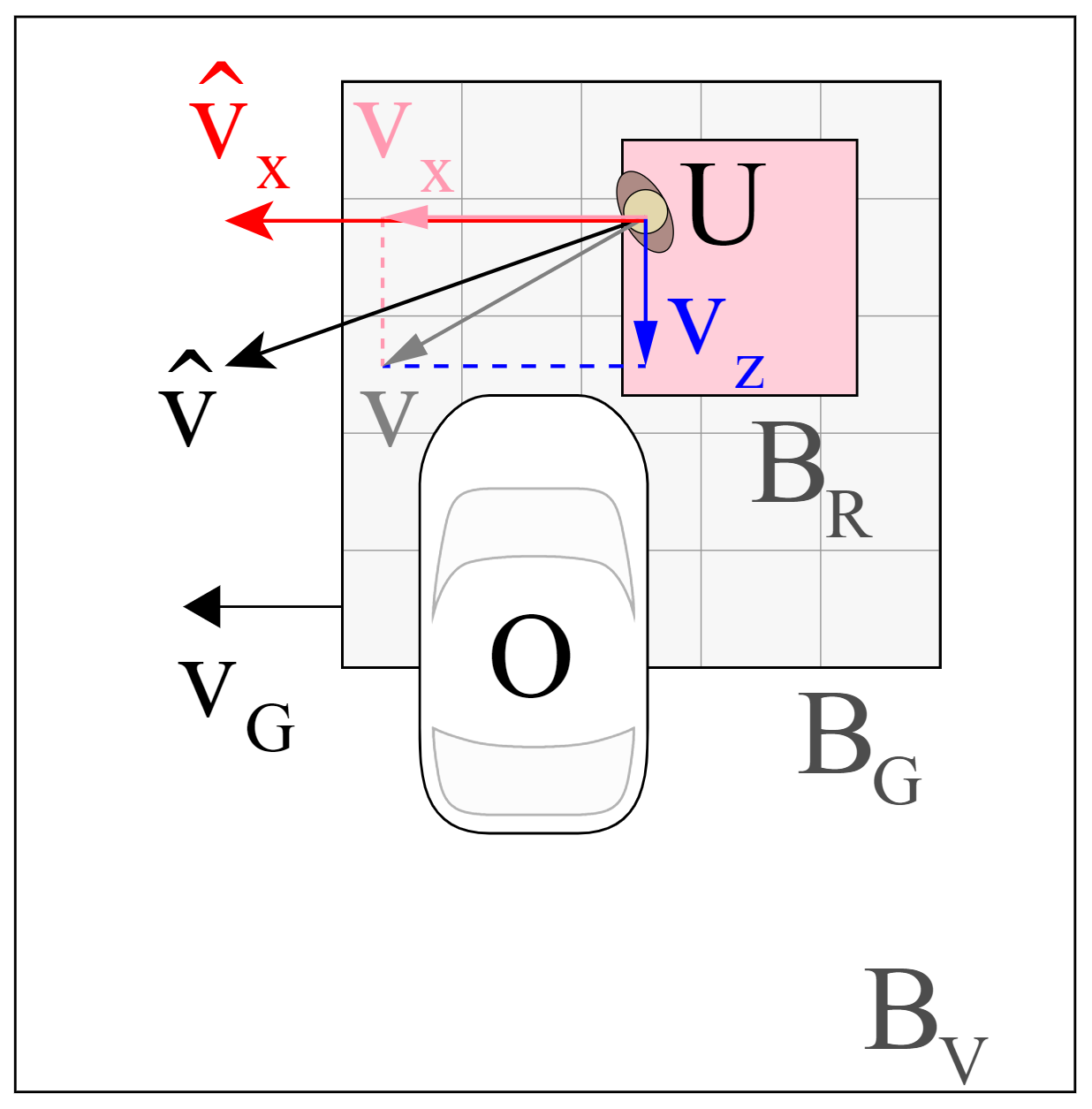}
    \caption{Example of rectangular movement gain where the user tries to move beyond an edge that is perpendicular to the $x$-axis. Thus, the gain is applied to the vector component $V_{_x}$ of real-world velocity $V$, which results in virtual velocity $\hat{V}$ with a vector component $\hat{V}_{_x}$, so that $V_{_G}=\hat{V}-V$.}
    \label{fig:rectgain}
\end{figure}

In Figure \ref{fig:rectgain}, the traversable virtual area is marked by boundary $B_{_V}$ and it contains the object of interest $O$. No movement gain is applied when the user $U$ moves within the rectangle $B_{_R}$. When the user attempts to move outside the edges of the rectangle with real world velocity $V$, then the virtual movement is amplified by using a gain constant to multiply the user's velocity components that are on the floor plane and perpendicular to the edges in question. This results in movement gain $V_{_G}$ and a total virtual velocity $\hat{V}$. Simultaneously, the rectangle $B_{_R}$ is shifted in the movement-gain direction so that the shifted edges will intersect the $(x,z)$ coordinates of the user's new position. In other words, the rectangle is dragged with the user whenever they try to step beyond its edges. The boundary $B_{_G}$ in Figure \ref{fig:rectgain} represents the physical play area. If we want to visualize how the physical play area boundary moves with the user in virtual world coordinates, we need to apply the gain velocity $V_{_G}$ to $B_{_G}$.

This technique is directly useful for any similar situation where there is an object of interest in the centre of the virtual environment. It might be compared to techniques that amplify movement at different scales (e.g., \cite{williams_updating_2006}). A general idea for exploration might be scaling techniques that exploit more complex relationships between virtual space and the available tracking space.

\subsection{Task-Oriented Teleportation}

\begin{figure}[h]
    \centering
    \includegraphics[width=0.5\linewidth]{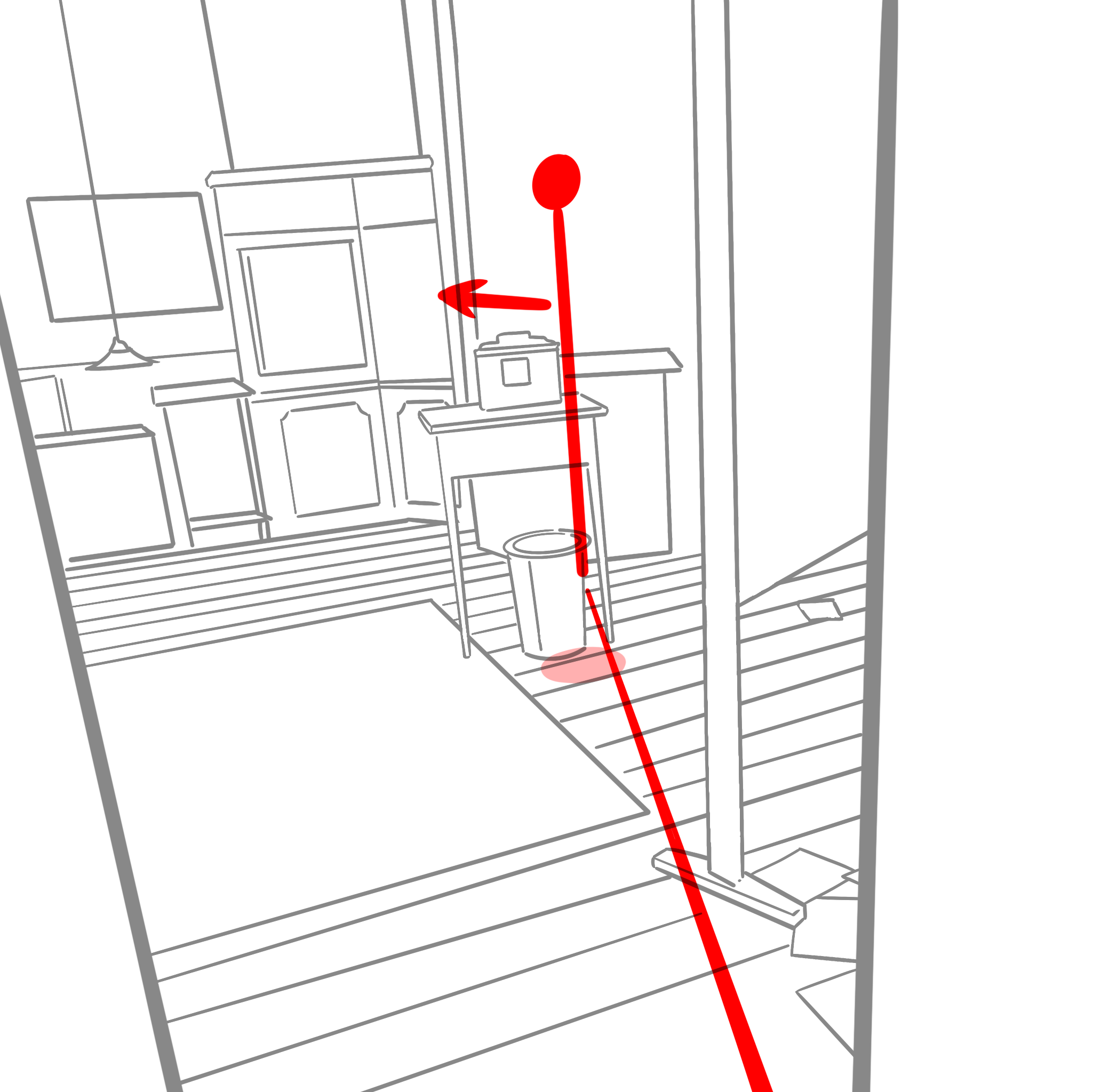}
    \caption{Task-oriented teleportation, in which a user can only move to fixed points with locked angles in order to perform certain actions, such as \textit{The Room VR}. The red arrow indicates the forward facing direction of the user after the teleport to the point indicated by the stick and teleport ray.
    \label{fig:taskjump}}
\end{figure}

Teleportation as a general approach has many advantages, as previously detailed. Designers have also come up with variations on the traditional location teleportation, in order to better suit their particular applications. One such variation is to constrain teleportation destinations to specific jump points (Figure~\ref{fig:taskjump}). This underscores another important point often missed in research studies, which is the important trade-off between precision and fun. While giving players complete control over their movements might seem like a good idea, it often imposes unnecessarily stringent precision requirements, leading to frustration and a drop in user enjoyment. Jump points can allow players to focus more on the core game mechanic, and fight less with the control scheme, as evidenced by \textit{\textbf{The Room VR}} experience. An early implementation of a similar scheme was the \textbf{\textit{Bullet Train}} demonstration, where not only was teleportation constrained to specific useful positions, but the rotation of the user was predetermined. This was due to this demonstration being available for the Oculus Rift DK2 which was constrained in its rotational tracking. In \textit{\textbf{The Room VR}}, there is a similar rotation on teleport, but it is more subtle; it is enabled only when the participant chooses a seated mode of operation, given that a seated user cannot turn all the way around. 

While it appears to diminish user freedom, there is interesting work to be done on whether this has an impact on user preference, situational awareness, etc. It is not difficult to implement within a scene, and it might be possible to analyse scenes to find teleportation locations.

\subsection{Rotation Teleportation}

\begin{figure}[h]
    \centering
    \includegraphics[width=0.5\linewidth]{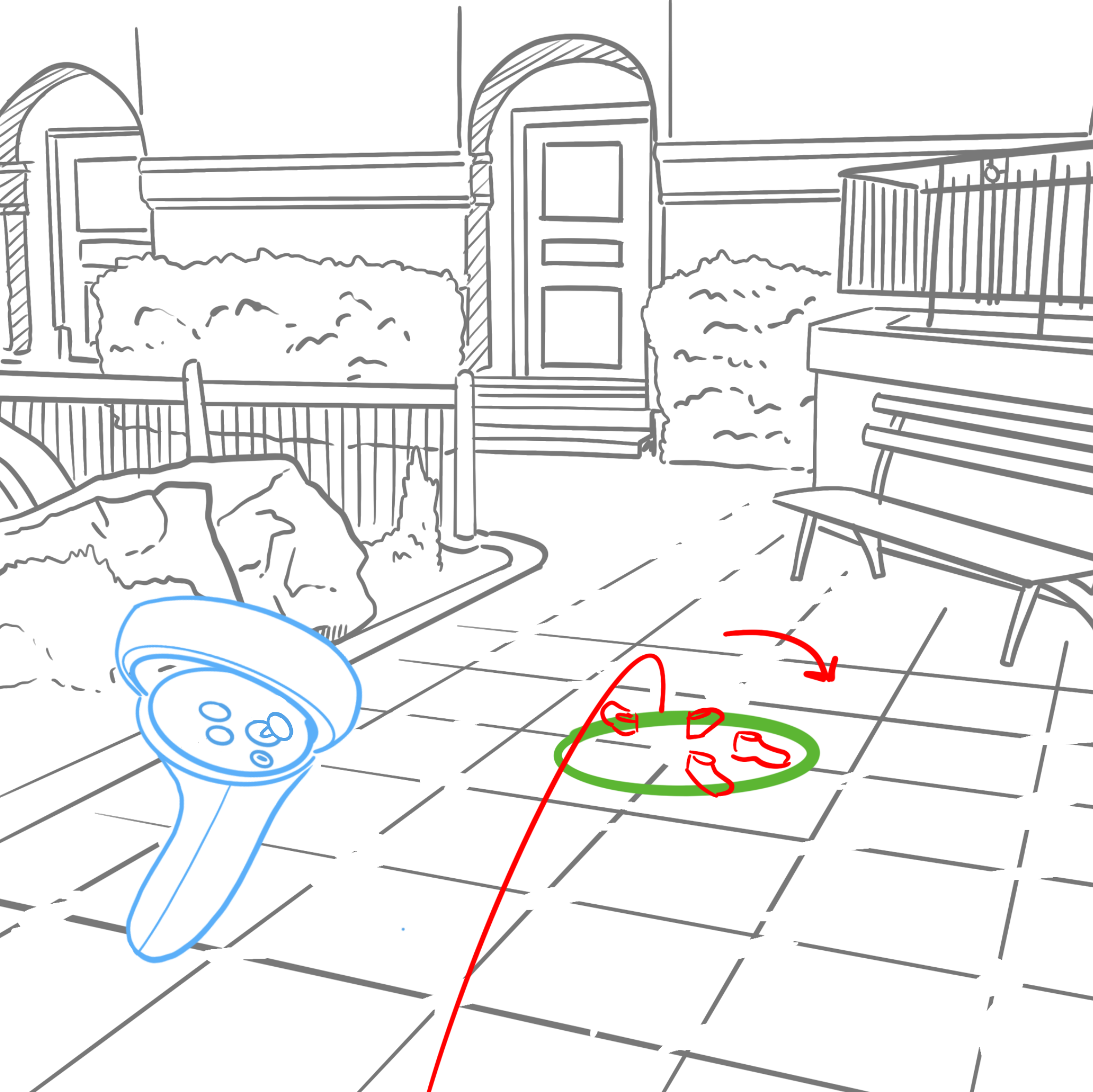}
    \caption{Rotational teleportation, whereby a user can select the direction they will face by rotating the joystick around the Roll axis during teleportation, as shown in \textit{Half Life: Alyx} }
    \label{fig:orientedjump}
\end{figure}

Another variation on teleportation includes support for freely indicating the desired orientation of the player once arriving at the designated jump point. This indication of the orientation is normally done with a visual cue, such as the footprints in \textit{\textbf{Half Life: Alyx}} (Figure~\ref{fig:orientedjump}), which rotate with the rotation of the controller about the Roll axis. This technique can be slower than the simple Point and Teleport technique, but does provide more control.

This technique solves a few problems, but does require more experience to execute efficiently. It can be used as an extension to any teleportation technique, but certain users might prefer not to use it. It does support seated mode and constrained tracking modes. We would suggest that this is in the realm of options that might generally be offered to users if they are comfortable using it.

\subsection{Preview Teleportation}

Yet another variation on teleportation, used in the game \textit{\textbf{Budget Cuts}}, allows players to have a preview of the jump point location before actually teleporting. This can give players more information about possible destinations, and might save time if unnecessary teleportations can be avoided.

\begin{figure}[h]
    \centering
    \includegraphics[width=0.32\linewidth]{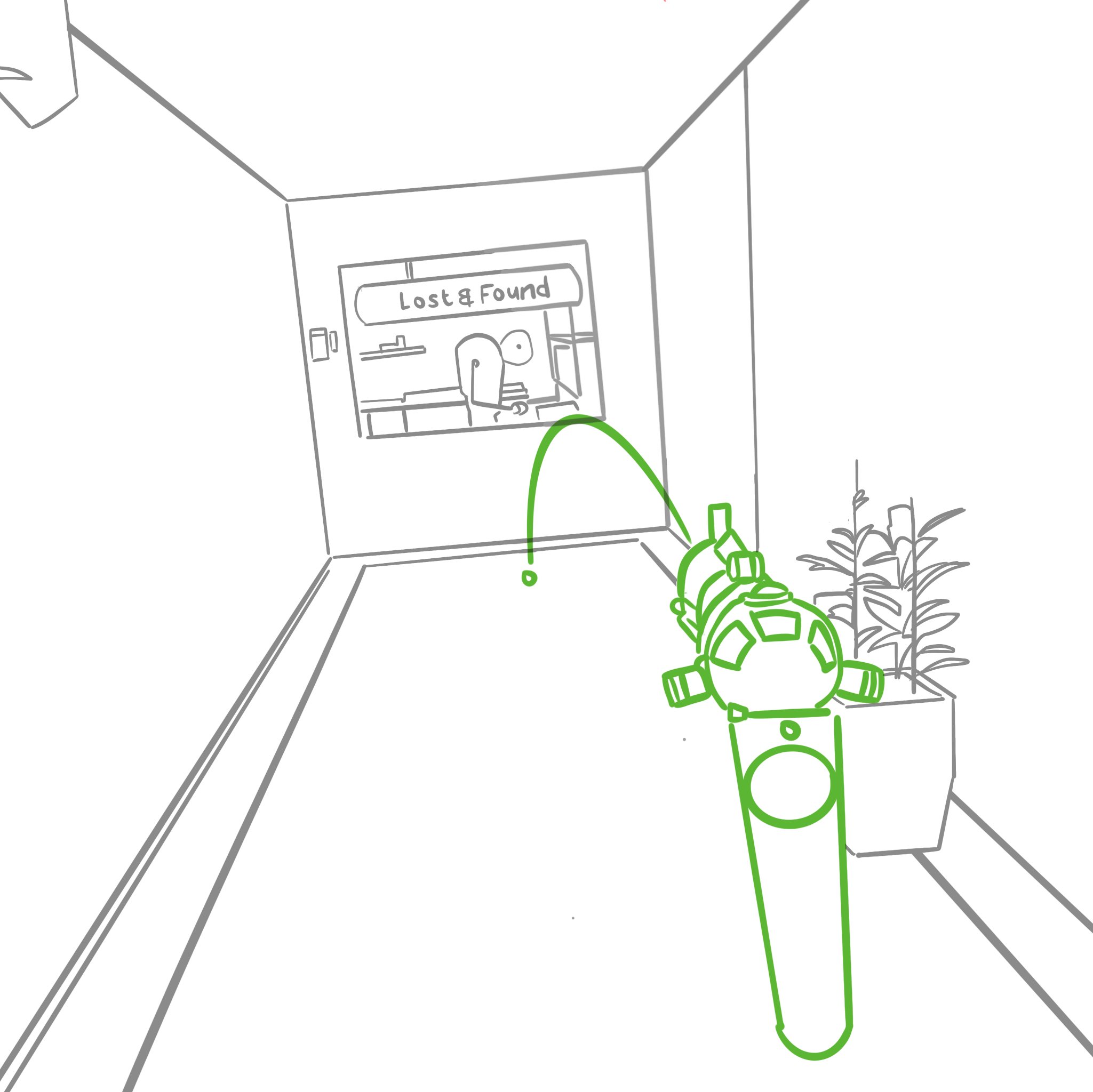}
    \includegraphics[width=0.32\linewidth]{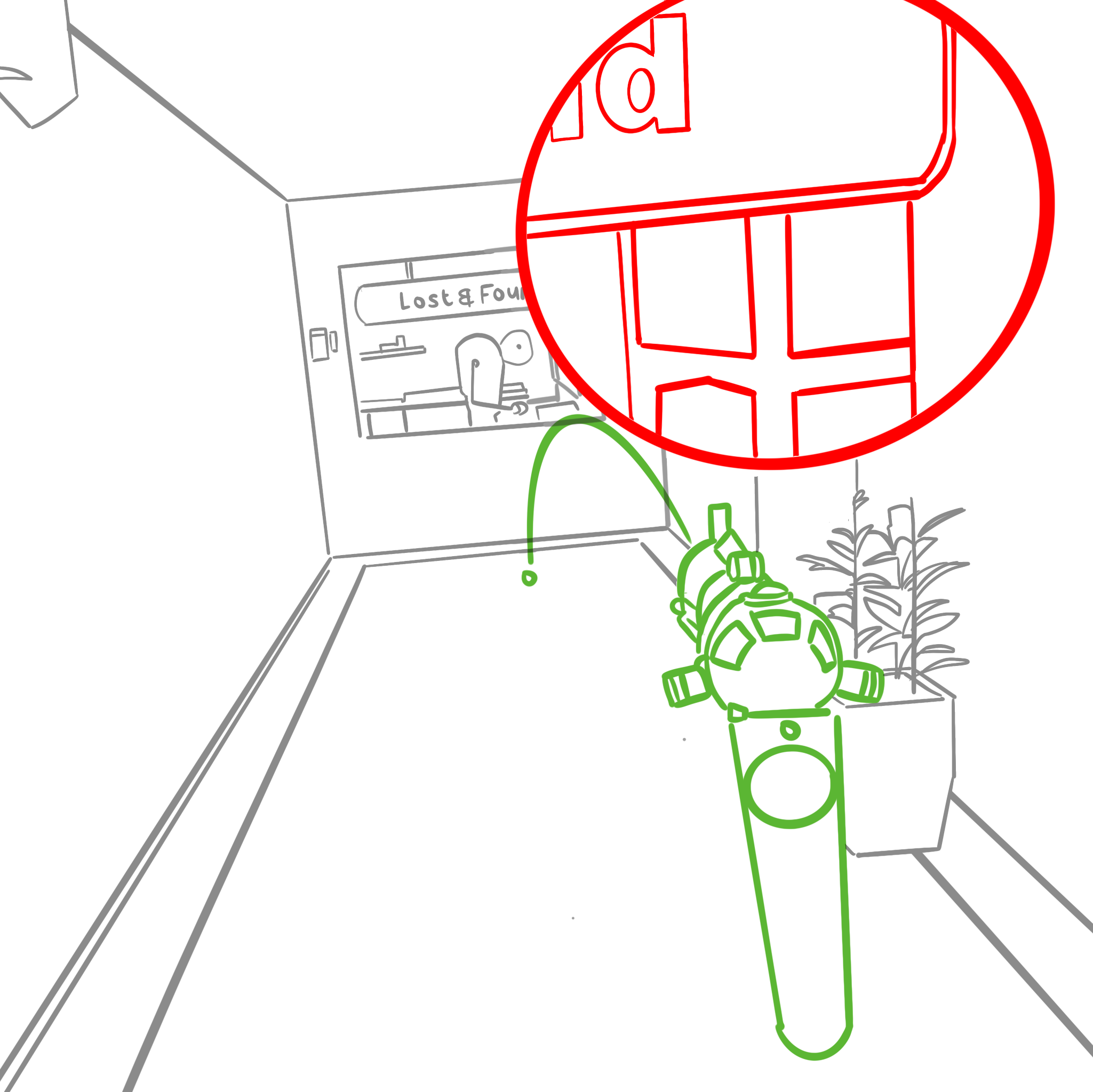}
    \includegraphics[width=0.32\linewidth]{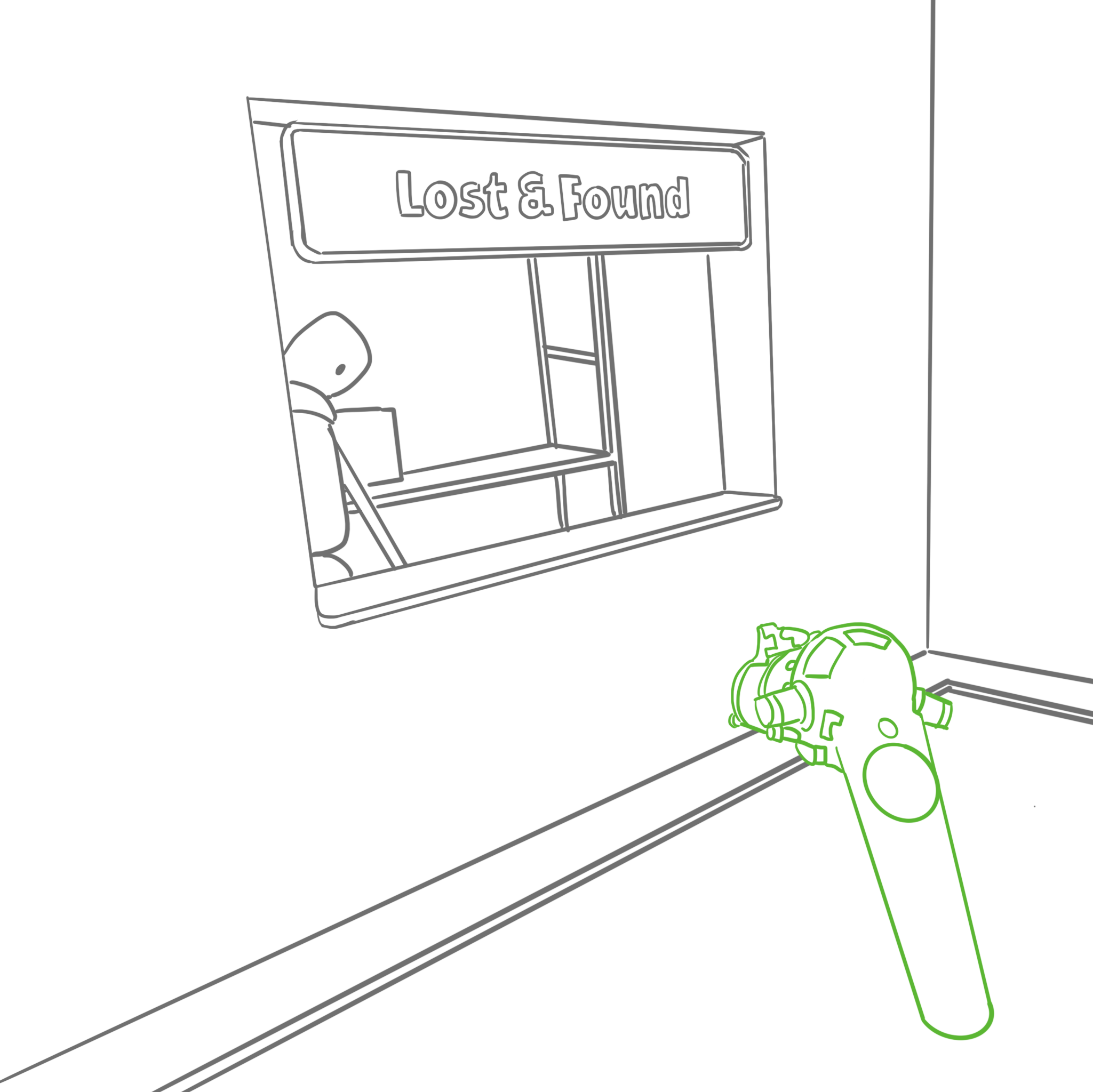}
    \caption{Translocator mechanic from \textit{Budget Cuts}, giving users a preview window POV from a future location before they choose to move there.}

    \label{fig:previewjump}
\end{figure}

This technique has general utility, and we suggest that it would be interesting to study its general application, especially when it comes to locomotion through complex environments or where locomotion accuracy is important. It might have interesting applications for situation awareness; this is certainly something that the game \textit{\textbf{Budget Cuts}} plays with as part of its design.

\section{System Control}
\label{sec:systemcontrol}

System control is a broad category covering virtual controls that are embedded within the virtual environment. As noted in Section \ref{sec:diegetic}, games have sometimes tried to adopt a diegetic approach where system controls are embedded into objects in the world. This is not always possible, especially if there are a large number of controls, or the controls do not have a logical diegetic representation. Thus, many current consumer VR applications include menu or control panels of some sort. Two common strategies are to include panels of controls suspended in space appearing to be in front of the user either at a comfortable distance (e.g., within arm's reach), or a few metres away. The former might be used by touching, the latter by ray selection, see Section \ref{sec:selection}. 

In this section, we discuss three patterns for system control, each of which has been explored in an academic context. It is interesting to see these designs re-emerge, and thus the main question is whether these can be generalised and recommended for certain classes of applications.

\subsection{Miniature Worlds}
\label{sec:wim}

The Worlds in Miniature (WIM) demonstration from Stoakley et al. \cite{stoakley_virtual_1995} is a seminal early demonstration of the potential for system control within an immersive environment. The user can interact with the larger environmental context, but also has a miniature copy of part of the environment that they can manipulate. This can bring distant objects within arm's reach and can solve some of the problems of selection and manipulation (c.f. Sections~\ref{sec:selection}~\&~\ref{sec:manipulation}). It is thus no surprise that the technique has been exploited in recent systems~\cite{danyluk_design_2021}. For example, Unity Technologies have demonstrated an immersive scene editor \textbf{\textit{EditorXR}} that has a MiniWorld component for showing parts of the world for editing. 

This technique has an obvious place in some applications, especially if they involve a lot of manipulation of object positions. We can, however, make a couple of observations. The first is that unlike most prior work on interaction that dealt primarily with human-scale scenes, a number of demonstrations, for example, \textbf{\textit{Lucky's Tale}} and \textbf{\textit{Allumette}}, use small-scale worlds, where the game or interaction takes place in a relatively small space, with much of the virtual world within arm's reach. This partly solves a problem with interaction at a distance, but also reduces the need for locomotion. Finally, we note that the WIM idea itself can be used as an interesting mechanism within game play, and we refer the interested reader to the game \textbf{\textit{A Fisherman's Tale}}.

\subsection{Body-Centred Menus}

\begin{figure}[h]
    \centering
    \includegraphics[width=0.5\linewidth]{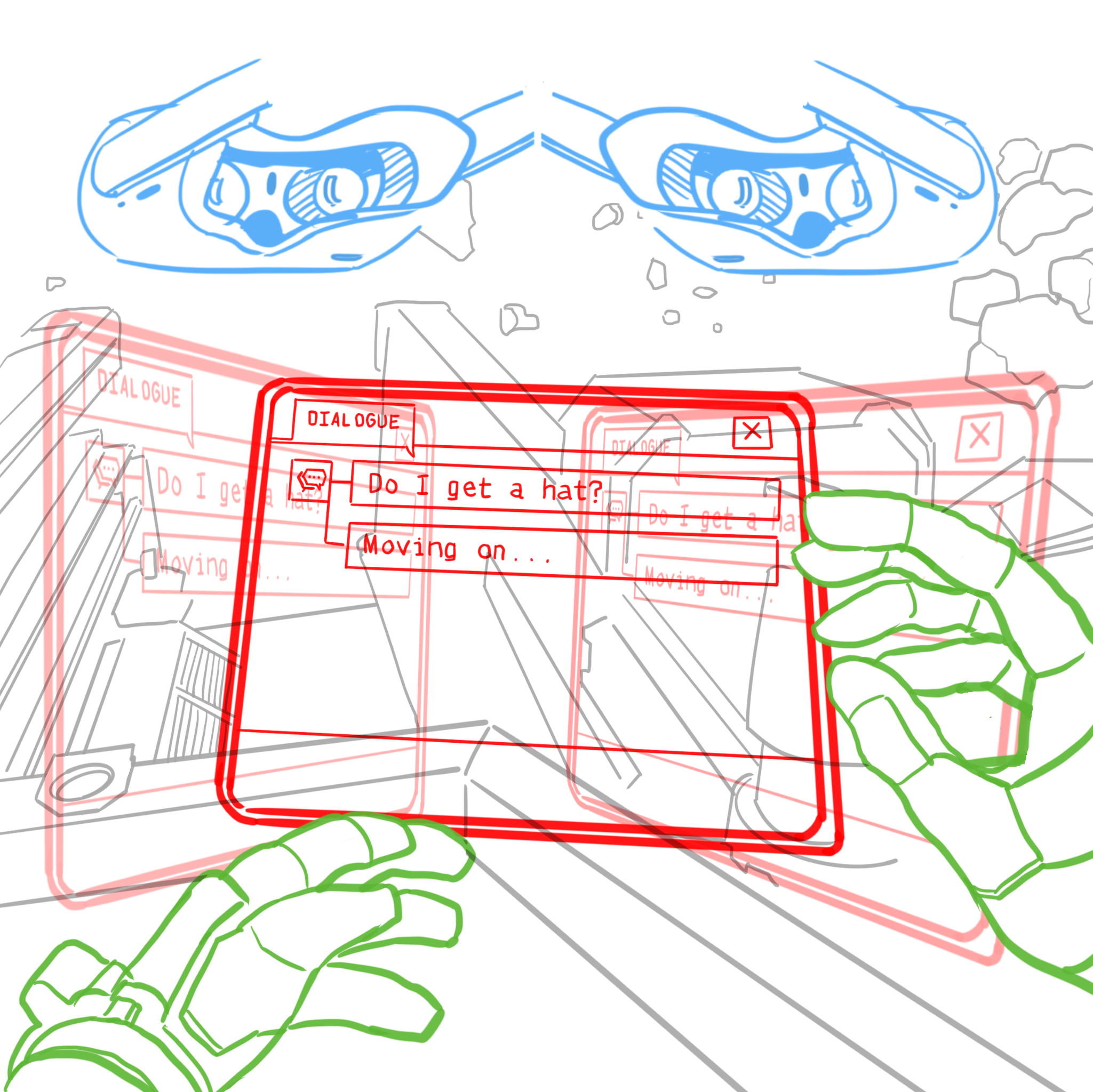}
    \caption{Body-centred menu from \textit{Lone Echo}.}
    \label{fig:menu}
\end{figure}

The idea of using the body of the user as a grounding for placing menus and controls has long been studied in immersive VR\cite{mine_moving_1997, Lindeman1999a, lediaeva_evaluation_2020}, and is a design choice that has been made in various games. \textbf{\textit{Lone Echo}} embeds controls into wearable controllers on the user's wrists (Figure~\ref{fig:menu}), fitting nicely with the futuristic theme of this game. It is perhaps interesting to note that a lot designs in this space are based on visions of what augmented reality/heads-up-displays might look like in the future. More grounded in reality is the technique in \textbf{\textit{RecRoom}}, where the main menu is activated by the user gesture of looking at the face of a wristwatch.

\subsection{In-Hand Menus}
\label{sec:inhand}

As with body-centred menus, in-hand menus are not a novel idea. They build on the natural ability for a user to use both hands to coordinate actions, with the dominant hand performing actions in a coordinate space determined by the non-dominant hand~\cite{buxton_study_1986, Lindeman1999a}. A body-centred menu also exploits this idea, but an in-hand menu might be constructed as a rather larger tool that can be manipulated by the non-dominant hand, with the dominant hand then selecting items. \textit{\textbf{TiltBrush}} is a good example of this; the in-hand tool is quite large and is manipulated by joystick or buttons on the controller that matches the hand in which it is held.

\section{Miscellaneous}
\label{sec:misc}

The following techniques fall into a class that is not well explored in prior work concerning 3DUI: external factors about the system or its impact on the user (c.f Section \ref{sec:dealing}). The techniques below deal with situations that might be outside the realm of interaction techniques themselves, e.g., concerned with user comfort or dealing with the equipment. Thus, they might be more generally used and again we flag that they deserve more study in our field, potentially as part of a broader effort to support practical use and accessibility (c.f. Section \ref{sec:access}).

\subsection{Independent Visual Background}

Many VR games include an Independent Visual Background (IVB) for the purpose of reducing visually-induced motion sickness. An IVB provides a visual reference frame that appears to be stationary with relation to the user's physical inertial environment \cite{duh2001independent}.

\begin{figure}[h]
    \centering
    \includegraphics[width=0.5\linewidth]{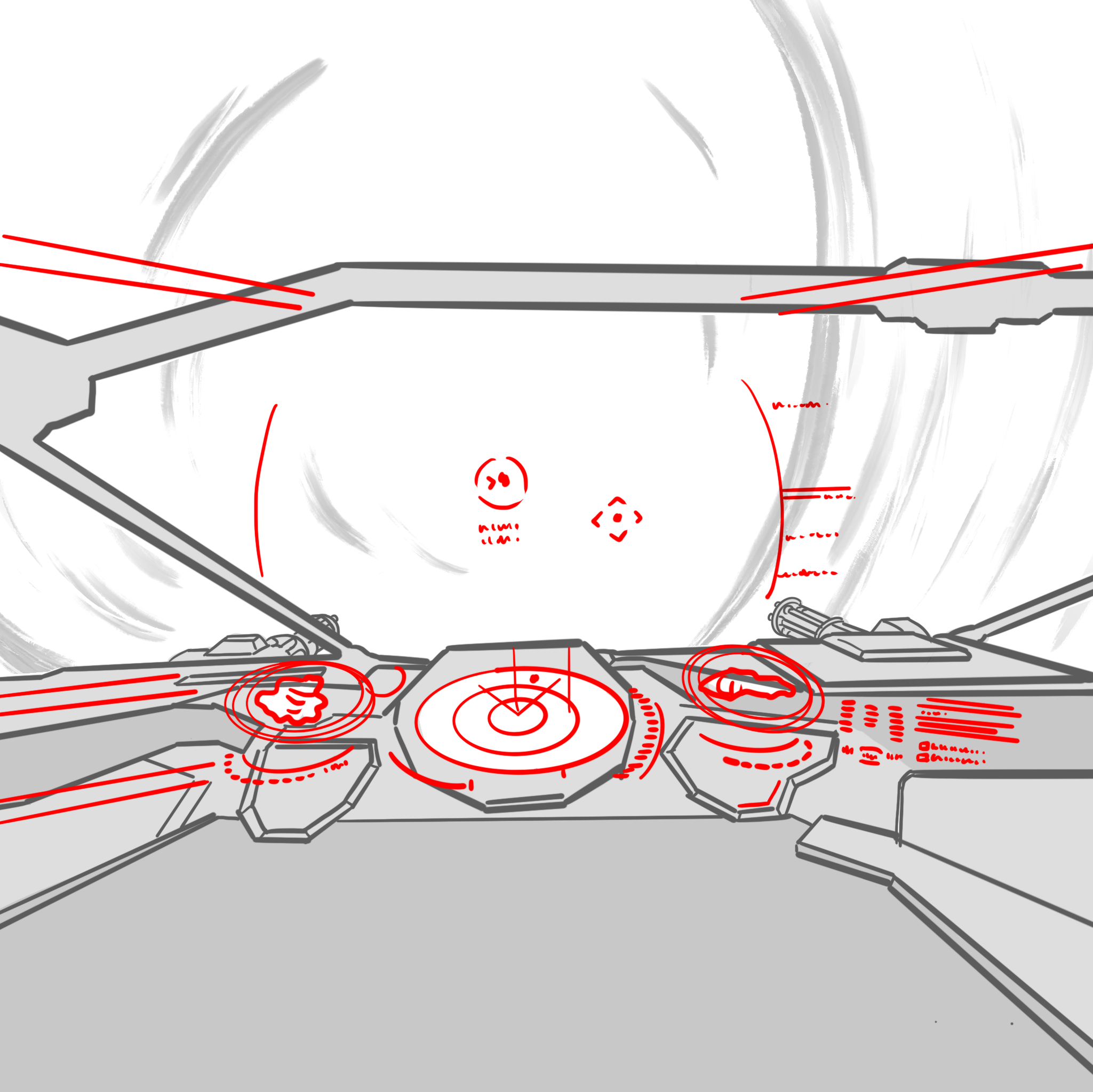}
    \caption{IVB cockpit from \textit{Elite Dangerous VR}.}
    \label{fig:IVB}
\end{figure}

For example, the game \textbf{\textit{Windlands 2}} features two IVBs that can be enabled by the user: a ``comfort cage'' that surrounds the player and floor markers. A similar IVB is available in the game \textbf{\textit{Espire 1}}, where the cage is visible only when the player moves virtually. This is also used in the rendition of speed lines in the game \textbf{\textit{Sprint Vector}}. Multiple IVBs are available for Unity and Unreal developers in the free \textit{\textbf{VR Tunnelling Pro}} plugin. Moreover, there are several VR games with cockpits that act as diegetic IVBs, e.g., \textbf{\textit{Elite Dangerous VR}} (Figure \ref{fig:IVB}), \textbf{\textit{Hover Junkers}} and \textit{\textbf{Vox Machinae}}.

Related to IVBs are visual reference frames that are locked to the user's head pose, which could be anything from heads-up displays to virtual noses \cite{wienrich2018virtual}. An example of the former can be found in the game \textbf{\textit{Lone Echo}}, while \textbf{\textit{Nintendo Labo's VR Blaster}} minigame features a head-mounted gun resembling a nose.

\subsection{Turn Signals for Cable Untangling}

HMDs that are connected via a cable to a PC or a console limit the user's range of movement. If the user keeps rotating several full turns in one direction, the cable can get wrapped around the user's body or chair, or other unwanted tangling might occur. To address this issue, some VR demos such as \textit{\textbf{Cosmic Wandering}} feature visual indicators that show the accumulated user rotation around the Up-axis, which implies cable twisting.

A more general solution is provided by \textbf{\textit{TurnSignal}}, which is a VR utility that can be used on top of any SteamVR compatible application. \textbf{\textit{TurnSignal}} displays a graphical icon on the floor plane that indicates the HMD cable winding direction and magnitude that has accumulated during the lifetime of running the utility. This helps the user to unwind the cable without the need to recall how many full turns they have taken in total.

\subsection{Space Adjustment}
\label{sec:space}

One of the main issues with accessibility (c.f. Section~\ref{sec:access}) is the height of the player. We have already noted some differences between play for seated and non-seated users. Users might also just be different heights and have different mobility. We have noted several games that provide some means to change scale (Section~\ref{sec:scale}). This is not completely generalisable, though, as sometimes the world needs to be experienced at a specific scale. One technique that can be used is simply to allow the user to move the world up and down to compensate for height to make the region that they need to move in more accessible. A good example is \textbf{\textit{Eleven Table Tennis}} where the table can be moved up and down. Another is the game \textbf{\textit{Brass Tactics}}, a game played over a large table top. They combine a locomotion-by-dragging technique, with a technique to move the table up and down by grabbing with both hands.

We can see that such a technique might be useful more generally. For example, sometimes the user wants to switch from sitting or standing, but the system has made an assumption about the reachable area at design time, or estimated user height at start time.



\section{Conclusion}
\label{sec:conclusion}

In this paper, we have described some examples of interesting designs from recent consumer VR games and demos that we feel deserve further study. There are now thousands of consumer experiences available, and to some extent, while there are some good design guidelines and toolkits, best practice is still emerging. 

We framed the discussion by reflecting on some of the differences between the demands of consumer VR games, compared to the different foci and stances of 3DUI research over the past three decades. We noted that reliability, fit to space, accessibility and fun all come to the forefront in consumer VR, whereas a lot of academic research has focused on task efficiency. We hope that the designs we highlight prompt new questions and reinvigorate discussion of the main measures and lenses through which we should evaluate interaction techniques for immersive VR.

The paper itself was also partly an experiment in crowd-sourcing. The motivation to look at crowd-sourcing was because of the sheer amount of consumer content to draw upon. Thus, we utilised a process of sourcing suggestions from email lists and other social media. Suggestions were collected on a Trello board that is publicly available at \url{https://trello.com/b/V1R2xM0u/consumer-vr3dui}. That board serves as an index to other resources, including a Slack channel, and even the source to this paper and its bibliography. 

Our primary reflection on the crowd-sourcing process was that engagement levels were very different. A small number of individuals amongst the authors had to push the process along, but suggestions and comments came from many quarters. Engagement was highest where there was a new announcement on some communication channel and there was already a growing list of suggestions. Thus, there may be some bias in that some of the original team seeded the list with their own suggestions. Further, due the nature of crowd-sourcing, individual contributors will probably highlight what is immediately obvious to them that is similar to what is already listed. Given that many of the contributors are academics, we might expect a slight bias towards work they have themselves contributed to. We don't think that these factors significantly biased the sample, as a broad range of consumer applications is discussed. If we were to run such a process in a similar (e.g., augmented reality) or narrower domain, the main suggestion we would have would be to seed the data collection with a broad set of examples. Examples of narrower domains that we haven't really been able to cover in this paper are the issue of symbolic input; the role of hand-tracking and full-body tracking;  and speech and gesture interfaces. The second suggestion is to make it obvious what comprises a contribution (e.g., see the discussion on Trello about what we mean by a ``generalisable technique'' rather than just an ``interesting design''). 

Finally, one hope we have is that this paper prompts more interaction between researchers and application developers. We encourage readers to add more materials to the Trello, board even if it is just a suggestion of an application to look at. There are already many similar examples to the ones we have used in the paper on the board, and there are more examples of applications that use the techniques we have discussed. We hope this list grows as more people engage in this exciting area at the intersection of academia and industry.

\section*{Products and Demonstrations Cited}

All applications, demonstrations, peripherals and videos cited in the document are listed by title rather than author. All were validated June 3rd 2021. Please refer to the Trello board for more examples.

\begin{itemize}[noitemsep,nolistsep]
\item Allumette by Penrose Studios \newline \url{https://store.steampowered.com/app/460850/Allumette/}
\item Anyland by Scott Lowe \& Philipp Lenssen \newline \url{http://anyland.com/} 
\item  Audi AG Virtual Reality Experience by Re'flekt GmbH \newline \url{https://www.youtube.com/watch?v=At_Zac4Xezw} and \newline
\url{https://www.re-flekt.com/portfolio-item/audi-virtual-reality-experience}
    \item Beat Saber by Beat Games \newline \url{https://beatsaber.com/}
    \item Brass Tactics by Hidden Path Entertainment \newline \url{https://www.hiddenpath.com/game/brass-tactics/}
    \item Bullet Train by Epic Games \newline \url{https://www.unrealengine.com/en-US/bullet-train}
    \item Cosmic Wandering VR by Punchey \newline \url{https://punchey.itch.io/cosmic-wandering}
    \item Cyberith Virtualizer from Cyberith GmbH \newline \url{https://www.cyberith.com/}
    \item Eagle Flight by Ubisoft \newline \url{https://www.ubisoft.com/en-gb/game/eagle-flight/}
    \item EditorXR from Unity Technologies \newline \url{https://github.com/Unity-Technologies/EditorXR}
    \item Eleven Table Tennis  by For Fun Games LLC \newline \url{https://elevenvr.com/}
    \item Elite Dangerous from Frontier Developments plc \newline \url{https://www.elitedangerous.com/}
    \item Espire 1 by Digital Lode \newline \url{https://espire1.com/}
    \item Fisherman's Tale by InnerspaceVR \newline \url{https://afishermanstale-game.com/}
    \item Gadgeteer by Metanaut Labs Inc \newline\url{https://gadgeteergame.com/}
    \item GingerVR by Samuel Ang  and John Quarles \newline \url{https://github.com/angsamuel/GingerVR}
    \item I Expect You to Die by Schell Games \newline \url{https://iexpectyoutodie.schellgames.com/}
    \item Hover Junckers by Stress Level Zero, LLC \newline \url{http://www.hoverjunkers.com/}
    \item Job Simulator by Owlchemy Labs \newline\url{https://jobsimulatorgame.com/}
    \item Lands End by ustwo games limited \newline \url{https://www.ustwogames.co.uk/games/lands-end}
    \item Lone Echo by Ready at Dawn Studios \newline \url{https://www.oculus.com/lone-echo/}
    \item Lucky's Tale by Playful \newline \url{https://www.oculus.com/experiences/rift/909129545868758}
    \item Manifest 99 by Project Flight School, LLC \newline \url{https://www.manifest99.com/}
    \item NeosVR by Solirax CoreDev s.r.o. \newline \url{https://neos.com/} 
    \item Nintendo Labo's VR Blaster by Nintendo \newline \url{https://www.nintendo.com/products/detail/labo-vr-kit-starter-set/}
    \item Quill by Facebook Technologies LLC \newline \url{https://quill.fb.com/}
    \item Rec Room by Rec Room Inc. \newline\url{https://rec.net/}
    \item ROVR from Wizdish Ltd \newline\url{https://rovr.systems/}
    \item The Lab by Valve \newline\url{https://store.steampowered.com/app/450390/The_Lab/}
    \item Spring Vector by Survios, Inc \newline\url{https://survios.com/sprintvector/}
    \item The Spy Who Shrunk Me VR by Catland \newline \url{https://store.steampowered.com/app/754850/The_Spy_Who_Shrunk_Me/}
    \item TiltBrush by Google \newline\url{https://www.tiltbrush.com/}
    \item TurnSignal by Benjamin McLean \newline \url{https://github.com/benotter/TurnSignal}
    \item Unseen Diplomacy by Triangular Pixels \newline\url{https://unseendiplomacy.com/}
    \item Vox Machinae by Space Bullet Dynamics Corporation \newline \url{http://www.voxmachinae.com/}
    \item VR Tunnelling Pro by Luke Thompson \newline \url{https://github.com/sigtrapgames/VrTunnellingPro-Unity} 
    \item VRChat by VRChat \newline \url{https://hello.vrchat.com/}
    \item VRZ: Torment by StormBringer Studios \newline \url{https://www.vrzgame.com/}
    \item WalkinVR by 2MW \newline\url{https://www.walkinvrdriver.com/}
    \item Wendy by Dmitry Kurilchenko and Miles Russo \newline\url{https://vrjam.devpost.com/submissions/36264-wendy}
    \item Windlands 2 by Psytec Games Ltd \newline \url{http://www.windlands.com/}
\end{itemize}


\bibliographystyle{unsrt}  
\bibliography{references}  

\end{document}